\documentclass[aps, prc, floatfix, 10pt, showkeys, showpacs, twocolumn]{revtex4-1}
\usepackage{color}
\usepackage{graphicx}
\usepackage{amsmath, amssymb,bm}
\usepackage{float}
\usepackage{array}

\begin{document}
\title{Determining Fundamental Properties of Matter Created in Ultrarelativistic Heavy-Ion Collisions}
\author{J.Novak}
\author{K.Novak}
\author{S.Pratt}
\author{J.Vredevoogd}
\affiliation{Department of Physics and Astronomy and National Superconducting Cyclotron Laboratory,
Michigan State University\\
East Lansing, Michigan 48824, USA}
\author{C.E. Coleman-Smith}
\affiliation{Department of Physics, Duke University\\
Durham, North Carolina 27708, USA}
\author{R.L. Wolpert}
\affiliation{Department of Statistical Science, Duke University\\
Durham, North Carolina 27708, USA}
\date{\today}
\begin{abstract}
  Posterior distributions for physical parameters describing relativistic heavy-ion
  collisions, such as the viscosity of the quark-gluon plasma, are extracted through a
  comparison of hydrodynamic-based transport models to experimental results from 100$A$
  GeV + 100$A$ GeV Au+Au collisions at the Relativistic Heavy Ion Collider (RHIC). By
  simultaneously varying six parameters and by evaluating several classes of observables,
  we are able to explore the complex intertwined dependencies of observables on model
  parameters. The methods proved a full multi-dimensional posterior distribution for the model
  output, including a range of acceptable values for each parameter, and reveal correlations between
  them. The breadth of observables and the number of parameters considered here go
  beyond previous studies in this field.  The statistical tools, which are based upon
  Gaussian Process emulators, are tested in detail and should be extendable to larger data
  sets and a higher number of parameters.
\end{abstract}



\maketitle

\section{Introduction}
Relativistic heavy ion collisions provide the means to study both the novel properties of
the quark gluon plasma and the fascinating nature of how it is created and
evolves. Unfortunately, experimental procedure is confined to measurements of the asymptotic
momenta of the particles comprising the collision's debris. Addressing the
fundamental questions concerning the properties of super-hadronic matter and the
collision's evolution inherently depends on large-scale multistage transport models. Such
models have improved significantly in recent years, and now typically combine viscous
hydrodynamic treatments for the evolution of the semi-thermalized quark-gluon plasma
($\sim$1-7 fm/$c$), and microscopic hadronic simulations to describe the dissolution and
breakup of the produced hadrons ($\sim$7-20 fm/$c$). For the first fm/$c$ of the
collision, when the system is too far from equilibrium for even a viscous hydrodynamic
treatment, quantitative modeling carries large uncertainties. If the profile and flow
of the matter being fed into the hydrodynamic treatment could be determined
phenomenologically, it would be invaluable in understanding how QCD saturation phenomena
affect the initial energy deposition and thermalization.

The data sets from the Relativistic Heavy Ion Collider (RHIC) and from the heavy ion
program at the Large Hadron Collider (LHC) are immense. The heterogenous nature of the
data, along with the strong interdependence of disparate observables with respect to basic
model parameters, makes any interpretation of the data challenging. The phenomenology of
heavy ion collisions has progressed despite these difficulties, primarily by identifying
the principal connections between model parameters and observables. For example, it is
well understood that the shear viscosity of the quark-gluon plasma strongly affects the
observed anisotropic flow coefficients. In an early analysis
\cite{Romatschke:2007mq}, the viscosity was adjusted until one finds a satisfactory fit
with the anisotropic flow coefficient $v_2$. The shortcoming of such an approach is that
several other unknown parameters, such as the spatial anisotropy of the initial state
\cite{Drescher:2007cd}, also affect $v_2$. In turn, each of these parameters also affects
numerous other observables. Similar approaches with more advanced models
\cite{Soltz:2012rk,Heinz:2011kt,Shen:2011zc,Shen:2011eg,Song:2011qa,Song:2010aq,Song:2010mg,Shen:2010uy,Song:2009rh}
have considered the variation of several parameters, and also
the effects of such parameters on spectra. However, due to the numerical costs
of running the models, these approaches have been unable to consider the simultaneous
variation of more than two or three parameters, or to consider a wider range of
experimental observables. These limitations compromise both the rigor and completeness of
the effort. A more rigorous and complete approach would be to perform a Markov-Chain Monte Carlo (MCMC) exploration of parameter space. MCMC calculations involve performing weighted random walks through parameter space, and producing a sampling of parameters weighted by the statistical likelihood. Such approaches involve running the model at hundreds of thousands, or even millions, of points in parameter space. For each point, the entire model would need to be run with sufficient statistics to compare to data. This would be untenable for models that require on the order of one day of CPU time to perform a calculation for a single point.

Other fields of science face similar challenges. A notable case is the extraction of
cosmological parameters from observations of fluctuations of the cosmic microwave
background \cite{Schneider:2008zf,Habib:2007ca}. Here the parameters are some of the most
fundamental in nature, such as the densities of dark matter and of dark energy. To
overcome the limitations of running the model a large number of times, a surrogate model
(a.k.a. an emulator) was developed to stand in for the true computer code. Rather than
re-running the full cosmological evolution model during the exploration of the  parameter
space, one runs the full model at only $\sim 100-200$ points in parameter space, carefully
chosen to best fill the overall space. A surrogate model was constructed that effectively
interpolated from the finite set of observations of the full model. The emulator was then
substituted for the full model for the MCMC exploration of parameter space. Similar ideas have
been applied to the field of galaxy formation \cite{Gomez:2012ak}. Here, we report first
results for a large scale surrogate-model-based statistical analysis of heavy-ion
collision data. A small scale application of these ideas was discussed in \cite{Petersen:2010zt}.

For this first effort, only a small subset of possible data will be considered, that
coming from 100$A$ GeV +100$A$ GeV Au+Au collisions at RHIC. Spectra for pions, kaons and
protons will be considered, along with the elliptic flow observable $v_2$ measured for
pions, and femtoscopic source radii from two-pion correlations. The motivation for first
considering soft observables is two-fold. First, they are the most sensitive to the model
parameters related to the bulk properties of matter, and secondly, the dependencies are
highly intertwined. During the last two years, the data set for relativistic heavy ion
collisions has greatly expanded with the beam-energy scan at RHIC, and with the inaugural
heavy-ion run at the LHC. The set is rapidly growing as data is analyzed
from Cu+Au and from U+U runs at RHIC. Ultimately, one may wish to incorporate other
observables, such as dilepton emission, higher flow moments, species-dependent flow, or long-range correlations,
once the theoretical treatments become more standardized and robust. The methods described
here should scale well with increasingly large data sets, and incorporating additional
observables into the analysis should be tractable.

On the theory side, numerous parameters factor into models of heavy-ion
collisions. Several of these parameters are needed to describe the initial energy density
and flow profiles that comprise the initial state of the hydrodynamic evolution. Other parameters
describe the bulk properties of super-hadronic matter, such as the equation of state and
viscosities. Still other parameters could describe out-of-equilibrium behavior such as
chemical abundances of various quark species. Since this is the first application of emulators for describing heterogenous data in this field, only a half dozen parameters will be considered for this study. Four of the parameters describe the initial state for the
hydrodynamic module, and two describe the shear viscosity and its energy dependence above
the transition temperature. The equation of state from lattice calculations
\cite{Borsanyi:2010cj,Bazavov:2009zn} will be assumed to be correct. In a future study, that too will be
parameterized to learn to what extent the equation of state is constrained
experimentally. Hadronization will be assumed to produce a chemically-equilibrated
hadronic gas when the temperature reaches 170 MeV. In the future, this assumption will
also be relaxed and the away-from-equilibrium properties of these hadrons will be
parameterized. Additionally, one should expect a non-negligible bulk viscosity in the
transition region \cite{Paech:2006st,Karsch:2007jc}. However, due to some numerical
instabilities with handling bulk viscosity, it will be set to zero for this
study. An advantage of surrogate model techniques is that they scale well with an
increasing number of parameters, and the efficiency of the methods are not greatly
diminished if several parameters have only marginal impact. We expect these methods to
continue to work even if we triple the number of parameters.

Details of the model and data used for the analysis are provided in the next two
sections. The theory of the model emulator is described in Sec. IV, with a test of the emulator
against a mock data set in the subsequent section. Results from an analysis of the real
data set are given in Sec. VII, while a summary and outlook comprise the final section.

\section{Modeling the Evolution at RHIC}
\label{sec:rhicmodel}

For this study, four elements are involved in the modeling.
\begin{enumerate}\itemsep=0pt
\item The pre-thermal, or stopping stage. Rather than dynamically solving for the evolution during this stage, we apply a parameterized description of the stress-energy tensor describing the state of the collision at a time of $\tau_0=0.8$ fm/$c$. Although sophisticated models of the initial state do exist, e.g. 
\cite{McLerran:1993ni,Kharzeev:2000ph, Kharzeev:2002ei, Dumitru:2007qr, Schenke:2012wb},
the large uncertainties and the lack of theoretical consensus dissuades one from picking any individual model.
\item The hydrodynamic stage lasts from 0.8 fm/$c$ until the system falls below a hadronization temperature of 170 MeV. Viscous hydrodynamics is justified for a strongly interacting system that is not too far from equilibrium, and is especially convenient for a system undergoing a transition in degrees of freedom, because the equations can be applicable even when there are no well-defined quasi-particles.
\item Once the density has fallen to the point that the evolution can be modeled as binary collisions of hadrons, we switch to a microscopic hadronic simulation, or cascade. The cascade is able to account for the loss of equilibrium between species, e.g., the protons and pions moving with different average velocities or having different kinetic temperatures. The cascade also handles disassociation seamlessly. 
\item Final-state particles are correlated at small relative momentum due to interactions in the asymptotic state. Assuming that interactions with third bodies are randomizing, one can calculate two-particle correlations given the source function, which describes the distribution of relative distances between two particles of similar velocities. Taking the source function from the information about last collisions in the cascade, and convoluting with the square of the known outgoing relative two-particle wave functions, we calculate correlations, and from the correlations calculate effective Gaussian source radii which can be compared to those extracted from experimentally measured correlations functions.
\end{enumerate}

\subsection{Parameterizing the initial state}

Rather than applying one of the competing models for the initial state, a parameterized form is used for the initial energy-density and flow profiles. Three parameters describe the initial energy-density profile and one describes the flow profile. The first is a weight, $f_{\rm wn}$ between a wounded-nucleon profile and a saturation-based profile,
\begin{equation}
\label{eq:icweights}
\epsilon(x,y)=f_{\rm wn}\epsilon_{\rm wn}(x,y) + (1-f_{\rm wn})\epsilon_{\rm sat}(x,y).
\end{equation}
The wounded-nucleon profile 
\cite{Miller:2007ri}
and the saturation profiles are based on Glauber thickness functions which describe the projected areal densities of the incoming nucelei in a plane perpendicular to the beam axis,
\begin{equation}
T_{A,B}(x,y)=\int dz~ \rho_{A,B}(x,y,z),
\end{equation}
where $\rho_{A,B}$ are the baryon densities of the two nuclei given the impact parameters. The thickness functions have units of baryons per fm$^2$, and the energy densities have the form,
\begin{widetext}
\begin{eqnarray}
\label{eq:wn}
\epsilon_{\rm wn}(x,y)&=&
\frac{(dE/dy)_{pp}\sigma_{\rm nn}}{2\sigma_{\rm sat.}}
T_A(x,y)\left(1-\exp(-T_B(x,y)\sigma_{\rm sat})\right),\\
\nonumber
&&+\frac{(dE/dy)_{pp}\sigma_{\rm nn}}{2\sigma_{\rm sat.}}T_B(x,y)\left(1-\exp(-T_A(x,y)\sigma_{\rm sat})\right),\\
\label{eq:sat}
\epsilon_{\rm sat}(x,y)&=&\frac{(dE/dy)_{pp}\sigma_{\rm nn}}{\sigma_{\rm sat.}}
T_{\rm min}(x,y)\left(1-\exp(-T_{\rm max}(x,y)\sigma_{\rm sat})\right),\\
\nonumber
T_{\rm min}&\equiv&\frac{2T_AT_B}{T_A+T_B},~~T_{\rm max}\equiv(T_A+T_B)/2.
\end{eqnarray}
Here, the energy densities are per transverse area and per longitudinal rapidity, i.e., one would divide by the initial time $\tau_0$ to get energy per  fm$^3$. The three parameters are $f_{\rm wn}$, the saturation cross section $\sigma_{\rm sat}$, and the normalization $(dE/dy)_{pp}$. When two identical columns of nuclei collide, $T_A=T_B$, which leads to $\epsilon_{\rm wn}=\epsilon_{\rm sat}$. The quantity $\sigma_{\rm nn}$ is not an adjustable parameter, it is the known inelastic nucleon-nucleon cross section of 42 mb.

In the diffuse limit, where $T_A,T_B\rightarrow 0$, the energy density becomes $(dE/dy)_{pp}T_AT_B\sigma_{\rm nn}$, which is known as the binary collision limit. If one considers two diffuse nuclei colliding randomly over a large area $S$, one finds the average energy per area in either expression to be
\begin{equation}
\left\langle dE/d\eta\right\rangle=\frac{\sigma_{\rm nn}(dE/dy)_{pp}}{S}\int dxdy~T_A(x,y)\int dx'dy'~T_B(x',y')
=\frac{AB\sigma_{\rm nn}(dE/dy)_{pp}}{S}.
\end{equation}
\end{widetext}

The parameter $(dE/dy)_{pp}$ is the energy per unit rapidity of a $pp$ collision. Although that number is measured in the asymptotic limit, it might be different at the time hydrodynamics is initialized, $\tau_0=0.8$ fm/$c$. Since the model requires the energy density at $\tau_0$, the initial energy per unit rapidity becomes an extra parameter that is adjusted from 0.85 to 1.2 times the energy per rapidity of a $pp$ collision of \cite{Abelev:2008ab}.

The parameter $\sigma_{\rm sat}$ controls the scale for changing the behavior of $\epsilon_{\rm sat}$ from the binary collision limit where $\epsilon\sim T_AT_B$ to the saturated limit when $\epsilon\sim T_{\rm min}$. The change occurs for $T_{\rm max}\approx 1/\sigma_{\rm sat}$. The parameter $\sigma_{\rm sat}$ also changes the wounded nucleon scaling form from that of binary collisions to the saturated limit where it is proportional to $T_A+T_B$.

The wounded nucleon and saturation expressions differ when $T_a\ne T_b$. For the case where $\sigma_{\rm sat}T_a \gg 1$ and $\sigma_{\rm sat}T_b \ll1$, 
\begin{eqnarray}
\lim_{\substack{\sigma_{\rm sat}T_a\gg1\\ \sigma_{\rm sat}T_b\ll1}} \epsilon_{\rm wn} &=&\frac{(dE/dy)_{pp}\sigma_{nn}}{\sigma_{\rm sat}}T_a/2,\\
\nonumber
\lim_{\substack{\sigma_{\rm sat}T_a\gg1\\ \sigma_{\rm sat}T_b\ll1}} \epsilon_{\rm sat} &=&\frac{(dE/dy)_{pp}\sigma_{nn}}{\sigma_{\rm sat}}2T_b.
\end{eqnarray}

For a single nucleon, $\sigma_{\rm sat}T_b\ll1$, colliding onto a thick target, $\sigma_{\rm sat}T_a\gg1$, the energy density in the wounded nucleon expression continues to scale proportional to $T_a$. For example, colliding a single nucleon onto a target with $\sigma_{\rm sat}T_a=10^6$ would give nearly 1000 times the multiplicity for a collision with $\sigma_{\rm sat}T_a=1000$. In contrast, the saturation formula would give roughly the same energy density for both instances. It was shown in \cite{Drescher:2007cd} that differences such as these significantly affect the initial elliptic anisotropy, and therefore significantly affect the measured elliptic flow. This can be understood by considering the collision of two equal mass nuclei with an impact parameter in the $x$ direction. Along the $x=0$ line, both the wounded nucleon and saturation expressions give the same energy density. However, if one goes outward so that $x$ becomes sufficiently large that one is at the edge of one nucleus, while being near the center of the other nucleus, the wounded nucleon formula gives a significantly higher energy density. This gives a relatively lower elliptic anisotropy for the wounded-nucleon model, and results in lower elliptic flow for the wounded-nucleon form than for the saturation form.

The fourth varied parameter describes the initial transverse flow, i.e., the collective flow at $\tau_0=0.8$ fm/$x$. Initial flow has been found to significantly affect femtoscopic source sizes \cite{Pratt:2008qv} and elliptic flow \cite{JoshNewPaper}. In \cite{Vredevoogd:2008id,Vredevoogd:2009zu} it was shown that one can express the transverse flow as
\begin{equation}
\label{eq:univflow}
\frac{T_{0i}}{T_{00}}=\frac{- \partial_i T_{00}}{2T_{00}} \tau,
\end{equation}
given four conditions: (a) A traceless stress-energy tensor, (b) Lowest order in $\tau$, (c) Bjorken boost-invariance, (d) Anisotropy of the stress-energy tensor being independent of $x$ and $y$. The power of the parameterization is that in the high-energy limit one expects each of these conditions to be reasonably met. However, at finite energy and for higher orders in $\tau$, Eq. \ref{eq:univflow} can only serve as a guide to set a scale for the initial flow and cannot be trusted to better than a factor of two. For that reason, the initial flow is parameterized as a constant $F_{\rm flow}$ multiplied by the amount given in Eq. \ref{eq:univflow} for $T_{0i}/T_{00}$. The fraction $F_{\rm flow}$ was varied from 0.25 to 1.25. 

For this first study, the initial energy density profiles are calculated from the average areal densities of the incoming nuclei, and are smooth, as if many events from the same impact parameter were averaged together. This is known to be fairly unrealistic, and the shortcoming will be addressed in the future.

\subsection{Hydrodynamic Module}

Viscous hydrodynamics in an environment where there are no net conserved charges is based on local energy momentum conservation plus two assumptions. First, it is assumed that in the rest frame of the stress-energy tensor the effective pressure equals the equilibrated pressure,
\begin{equation}
\frac{T_{xx}+T_{yy}+T_{zz}}{3}=P(\epsilon),
\end{equation}
i.e. the bulk viscosity is assumed to be zero. Second, it is assumed that the remainder of the stress-energy tensor is sufficiently close its Navier Stokes value that its evolution can be described with Israel-Stewart equations of motion, which in the frame of the fluid becomes
\begin{eqnarray}
\label{eq:IS}
\pi_{ij}&\equiv&T_{ij}-\frac{1}{3}\delta_{ij}(T_{xx}+T_{yy}+T_{zz}),\\
\nonumber
\frac{d}{dt}\frac{\pi_{ij}}{\sigma(\epsilon)}&=&-\frac{1}{\sigma(\epsilon)\tau_{\rm IS}}\left(\pi_{ij}-\pi_{ij}^{\rm(NS)}\right),\\
\nonumber
\pi_{ij}^{NS}&=&-\eta\left(\partial_iv_j+\partial_jv_i-\frac{2}{3}\delta_{ij}\nabla\cdot{\bf v}\right).
\end{eqnarray}
The Israel-Stewart relaxation time was set to, $\tau_{\rm IS}=3\eta/sT$. The factor of three was chosen for being midway between the expectations for AdS/CFT theory \cite{Romatschke:2009im} and that of a Boltzmann gas of massless particles. For AdS/CFT, the factor would be replaced by $4(1-\ln 2)\approx 1.23$, whereas for an ideal gas one expects a factor of five. This can be seen from Kubo relations,
\begin{eqnarray}
\eta&=&\frac{1}{T}\int d^3r\int_0^\infty dt~\langle T_{xy}(0,0)T_{xy}(r,t)\rangle,\\
\nonumber
&=&\frac{\tau_{\rm IS}}{T}\int d^3r\langle T_{xy}(0,0)T_{xy}(r,t)\rangle,\\
\nonumber
&=&\frac{\tau_{\rm IS}}{T}\int d^3p~\frac{1}{(2\pi)^3}\frac{p_x^2p_y^2}{p^2}e^{-p/T}\\
&=&\frac{sT\tau_{\rm IS}}{5}.
\end{eqnarray}
Here, the first step comes from assuming the correlations die exponentially, which is the basic assumption of Israel-Stewart hydrodynamics, and the second step is based on the equal-time correlations coming only from a particle being correlated with itself, which is assumed in a Boltzmann picture. The final step simply involves performing the angular integrals and comparing the answer to the corresponding integral for $sT=\epsilon +P$. If results are shown to be sensitive to $\tau_{\rm IS}$, it should be treated as a parameter.

Once these conditions are met, applying the local conservation of energy and momentum,
\begin{equation}
\partial_\mu T^{\mu\nu}=0,
\end{equation}
determines the evolution of the stress-energy tensor.

At high energy density the first assumption, that $\sum_iT_{ii}=3P(\epsilon)$, can be met even if the system is far from chemical or kinetic equilibrium. For a gas of weakly interacting massless particles, or even for a region dominated by weakly interacting classical fields, the condition is met regardless of the configuration of either the particles or the fields. Once the fireball cools down near the transition region, and conformal invariance is lost, this assumption becomes questionable. The second assumption may be poorly met during the first 1-2 fm/$c$. However, the impact of changing the anisotropy of the stress-energy tensor at early times tends to be rather small \cite{Vredevoogd:2008id}.

The hydrodynamic module used here is built on an assumption of longitudinal boost-invariance which allows the calculations to become effectively two-dimensional before solving Israel-Stewart equations of motion. This approach has been applied by numerous research groups 
\cite{Muronga:2001zk, Dusling:2007gi, Song:2007ux, Luzum:2008cw, Niemi:2011ix}. 
The reduction of the dimensionality is justified to better than the five percent level \cite{Vredevoogd:2012ui}. The equation of state, $P(\epsilon)$, comes from lattice calculations of Wuppertal-Budapest group 
\cite{Borsanyi:2010cj} 
for temperatures above the hadronization temperature, and use a hadron-gas equation of state at lower temperatures. The equation of state for temperatures just above the hadronization temperature is slightly modified from the lattice values to match the hadron gas value at the hadronization threshold.

For temperatures above 170 MeV/$c$, the viscosity to entropy density ratio was described with two parameters,
\begin{equation}
\label{eq:alphadef}
\frac{\eta}{s}=\left.\frac{\eta}{s}\right|_{T_c}+\alpha\ln\left(\frac{T}{T_c}\right),
\end{equation}
where $T_c$ is assumed to be 170 MeV. The first parameter, $\eta/s|_{T_c}$, describes the viscosity just above the hadronization threshold, while the second parameter, $\alpha$, describes the temperature dependence. This parameterization is not particularly well motivated, but by varying the parameter $\alpha$ one can gain some insight into how sensitive results are to temperature dependence.

The hydrodynamic/cascade interface temperature was set at a temperature of 170 MeV. Calculations were also performed for a hadronization temperature of 155 MeV, but those calculations consistently over-predicted the flow, or equivalently, under-predicted the number of hadrons for a given amount of transverse energy. It is the authors' intention to perform a detailed study of the sensitivity to the equation of state and the details of hadronization in a separate paper. A summary of model parameters is provided in Table \ref{table:ICpars}.
\begin{table*}
\centering
\begin{tabular}{|c|>{\centering}p{0.75\textwidth}|c|}
\hline
parameter & description & range\\
\hline
$(dE/dy)_{pp}$ & The initial energy per rapidity in the diffuse limit compared to measured value in $pp$ collision & 0.85--1.2\\
\hline
$\sigma_{\rm sat}$ & This controls how saturation sets in as function of areal density of the target or projectile. In the wounded nucleon model it is assumed to be the free nucleon-nucleon cross section of 42 mb & 30 mb--50 mb\\
\hline
$f_{wn}$ & Determines the relative weight of the wounded-nucleon and saturation formulas for the initial energy density described in (\ref{eq:wn},~\ref{eq:sat}) & 0--1\\
\hline
$F_{\rm flow}$ & Describes  the strength of the initial flow as a fraction of the amount described in Eq.\ref{eq:univflow} & 0.25--1.25\\
\hline
$\eta/s|_{T_c}$ & Viscosity to entropy density ratio for $T=170$ MeV & 0 -- 0.5\\
\hline
$\alpha$ & Temperature dependence of $\eta/s$ for temperatures above 170 MeV/$c$, i.e., $\eta/s=\eta/s|_{T_c}+\alpha\ln(T/T_c)$ & 0 - 5\\
\hline
\end{tabular}
\caption{\label{table:ICpars}
Summary of model parameters. Six model parameters were varied. The first four describe the initial state being fed into the hydrodynamic module, and the last two describe the viscosity and its energy dependence.}
\end{table*}

\subsection{Hadronic Cascade}

The hydrodynamic module was run until all elements cooled below 170 MeV. During the hydrodynamic evolution, the properties of the 170 MeV hypersurface were recorded. This included the position and flow velocity at the boundaries, and the anisotropy of the stress-energy tensor. Hadrons were generated with a Monte Carlo procedure ensuring that all elements of the stress-energy tensor were continuous across the hyper-surface. The method \cite{Pratt:2010jt} assumes that all species have a single time relaxation scale independent of their momentum. Other approaches have considered the effect of adding a momentum or species dependence to the relaxation time \cite{Dusling:2007gi}, but because this study considers only particles with low to moderate $p_t$, and because the particles interact a few more times in the cascade module, the details of the algorithmic choice are not expected to matter, as long as the stress-energy tensor is continuous across the boundary.

A list of particles as produced in the interface was then fed into the cascade on an event-by-event basis. For these studies, 4000 cascade events were produced for each impact parameter. The cascade code was inspired by the physics of the hadronic module of URQMD \cite{URQMD}, but was significantly rewritten to improve speed, and is labeled B3D \cite{b3d}. Hadrons were assumed to collide through resonances with Breit-Wigner forms, plus a simple $s-$wave elastic cross section of 10 mb. The $s-$wave cross section was chosen independent of momentum and particle species. The resonances from the particle data book \cite{particledatabook} with masses less than 2.2 GeV/$c^2$ were all included. On average, particles collided roughtly twice after being generated from the hydrodynamic interface. Pions had fewer collisions on average, while protons had more. The collisions in the cascade mainly affected the spectra and $v_2$ of protons. There are numerous ways to improve the cascade, such as more realistic cross sections, consistent time delays in scattering processes, mean-field effects, and Bose effects for pions. However, given the rather modest impact of the cascade at high energy, it is not expected that the observables would change significantly.

The B3D code runs approximately two orders of magnitude faster than URQMD for the calculations used here. This is mainly due to two improvements: better handling of the linked lists used to track collisions, and adding cyclic boundary conditions so that boost-invariance could be efficiently incorporated. The majority of the numerical expense of the calculations came from the cascade, and improving the speed allowed a greater number of points in parameter space to be explored. 

The cascade ran until all collisions ceased. For each outgoing particle, the momentum, particle ID and the space-time coordinates of the last interaction were recorded. Since the reaction plane is known, it is straight-forward to calculate the azimuthal anisotropy factor $v_2=\langle\cos 2\phi\rangle$. Spectra are efficiently calculated given that the cyclic boundary conditions make it possible to use all the particles when calculating the spectra at zero rapidity.

\subsection{Femtoscopic Correlations}
\begin{widetext}
Two-particle correlations at small relative momentum provide femtoscopic information about the phase space distributions. This information is expressed through the Koonin formula \cite{Koonin:1977fh,Lisa:2005dd},
\begin{eqnarray}
C({\bf K}=({\bf p}_1+{\bf p}_2)/2,{\bf k}=({\bf p}_1-{\bf p}_2)/2)
&=&\int d^3r~S({\bf K},{\bf r})\left|\phi_{\bf k}({\bf r})\right|^2,\\
\nonumber
S({\bf K},{\bf r})&\equiv&\frac{\int d^3r_1d^3r_2~f({\bf K},{\bf r}_1)f({\bf K},{\bf r}_2)\delta({\bf r}-({\bf r}_1-{\bf r}_2))}
{\int d^3r_1d^3r_2~f({\bf K},{\bf r}_1)f({\bf K},{\bf r}_2)}.
\end{eqnarray}
\end{widetext}
Here, $\phi_{\bf q}({\bf r})$ is the outgoing two-particle wavefunction, $f({\bf p},{\bf r})$ is the phase space density in the asymptotic state, and $S({\bf K},{\bf r})$ describes the chance that two particles with the same asymptotic momentum ${\bf K}$ would be separated by ${\bf r}$ should they not interact. Correlations provide the means to determine the coordinate-space information of $S({\bf K},r)$ from the measured correlations, $C({\bf K},{\bf k})$. Through a fitting procedure, one can infer source radii which fit the shape of $S({\bf K},{\bf r})$ with Gaussian radii, i.e. $S({\bf K},{\bf r})\sim \exp\{-x^2/2R_{\rm out}^2-y^2/2R_{\rm side}^2-z^2/2R_{\rm long}^2\}$, where the ``outward'' direction is transverse to the beam and parallel to ${\bf K}$, the ``longitudinal'' direction is along the beam axis and the ``sideward'' direction is perpendicular to the other two. The source function $S({\bf K},{\bf r})$ depends on both spatial and temporal aspects of the emission. For instance, if the source is small but long-lived, the outgoing phase space cloud for pions with momentum ${\bf K}$ will be elongated along the direction of ${\bf K}$ due to some pions being emitted long before others. This would lead to the extracted radius $R_{\rm out}$ being much larger than $R_{\rm side}$. In contrast, the two radii tend to be quite similar if the expansion is highly explosive.

The source radii are typically extracted by experimental collaborations through fitting their measured correlations to expectations from Gaussian sources. Description of such analyses can be found in \cite{Lisa:2005dd}. For the model calculations correlation functions were calculated by first sampling $S({\bf K},{\bf r})$ then combining pairs of pions with similar momentum. Pions were divided into bins of 20 MeV/$c$ width in transverse momentum and in 15$^\circ$ bins in azimuthal angle, before pairing. Utilizing boost invariance, all the pions could be longitudinally boosted to a frame where the rapidity was zero. The space-time points at which particle's had their last interaction had been recorded along with their asymptotic momentum during the running of the B3D module. This allowed a list of ${\bf r}={\bf r}_1-{\bf r}_2$ to be constructed for each momentum bin. Correlation functions for each momentum bin were calculated by assuming a simplified wave function, $|\phi_{\bf q}({\bf r})|^2=1+\cos(2{\bf k}\cdot{\bf r})$. Gaussian source radii were then found by searching for radii that best reproduce the three-dimensional correlation functions calculated by the model. A fourth parameter, usually referred to as $\lambda$, was also varied to describe the fraction of particles that are correlated, since a good fraction of pairs are uncorrelated since one of the particles may have resulted from a decay, or even have been misidentified. Thus, rather than matching experimental and theoretical correlation functions, Gaussian radii were compared. The calculation of correlation functions and fitting was performed with the code base in CorAL \cite{coral}.
\section{Reduction of Experimental Data for Statistical Analysis}
\label{sec:rhicdata}

The heavy ion data sets from RHIC and from the Pb+Pb experiments at the LHC represent some
of the largest scientific data sets in existence. A principal motivation of this work is
to develop a statistical analysis that can be extended to large heterogenous data
sets. This would include data taken at multiple beam energies, with different
target-projectile combinations and with different detectors. The recent beam-energy scan
at RHIC and the inauguration of the LHC have increased the available data by more than an
order of magnitude as compared to the Au+Au collisions at 100$A$ GeV beams measured at
RHIC. Additionally, analyzed measurements of Cu+Cu, Cu+Au and U+U from RHIC will soon be
available. The data set from the one beam energy contains petabytes of
information. For this first study, we confine our analysis to this one data set, Au+Au at
100$A$ GeV + 100$A$ GeV. We further confine the analysis to a subset of soft physics
observables: spectra, elliptic anisotropy, and femtoscopic correlations. Only mid-rapidity
observables were considered. These are the observables most connected to the bulk dynamics
and to the bulk properties of matter, and are often referred to as ``soft
physics''. Several classes of observables are being ignored, e.g., jet quenching, long-range
fluctuations and correlations, dilepton and direct-photon measurements, and heavy
flavor. These observables are often labeled ``rare probes'' and their interpretation
largely factorizes out of the analysis of the soft observables being considered here. For
instance, although jet quenching depends on the energy density and bulk properties of the
quark gluon plasma, the soft physics observables being considered here are not
significantly affected by the mechanism for jet production. Further, the theory and
phenomenology governing these other classes of observables often carry large
uncertainties, not only in additional unknown parameters, but also in that they carry
questions concerning the choice of approach. Given the way that the physics from these
other classes of analyses factorize from the soft physics, and the lack of theoretical
consensus, the prudent course of action is to determine the bulk dynamics of the
system using the soft physics observables. Once the evolution of the system is determined,
with quantified uncertainties, one would have a validated basis from which to calculate
other classes of observables, such as rare probes.

Within the set of soft-physics observables, this first analysis is restricted to a
subset of the overall data. For spectra, we consider only pions, kaons and protons. It
would be straight-forward to consider strange baryon spectra, but due to large systematic
and statistical errors, they are unlikely to greatly affect the answer at the current
time. Additionally, because theoretical treatments away from mid-rapidity remain in an
immature stage, our analysis concerns only mid-rapidity observables. For angular
anisotropies, we consider only $v_2$ and ignore higher order anisotropies for $n>2$,
\begin{equation}
v_n\equiv\langle \cos(n\phi)\rangle,
\end{equation}
where $\phi$ is the angle of a particle relative to the reaction plane. Recent analyses of
$v_{n>2}$ suggest that the observables may even be more sensitive to the viscosity than
$v_2$ \cite{Schenke:2010rr,Schenke:2011bn,Alver:2010dn}. However, theoretical questions remain about
how to instantiate the event-by-event fluctuations which drive these higher-order
harmonics. This analysis only considers $v_2$ for pions. Although $v_2$ is measured for
kaons and protons, in order to compare to data, theoretical treatments would have to run
for tens of thousands of events for each impact parameter to get sufficient statistics for
kaons and protons. This analysis used 4000 events per impact parameter. Finally, the
femtoscopic analysis is confined to same-sign pions. Source sizes extracted from other
analyses carry significantly more uncertainty. RHIC data is recorded according to
centrality bins, e.g., top 5\%, top 10\%, 10-20\% $\ldots$. Bins are typically assigned
according to some measure of overall multiplicity. For instance, the 20-30\% bin
corresponds to those events with multiplicities that are lower than the top 20\% of events
and higher than the lower 70\% of events. The choice of bins varies between observables
and between collaborations. Since hydrodynamic treatments, especially the use of smooth initial conditions, becomes more questionable at low centrality, we decided to neglect the more peripheral collisions. Even though hydrodynamic models have been successful in fitting data for centralities up to 50\% \cite{Song:2011qa}, we have chosen to ignore
centralities greater than 30\%. For future analyses that include initial-state fluctuations, less central collisions should be included, especially since they might provide more sensitivity to the initial flow. This analysis was thus confined to two bins, 0-5\% and 20-30\%, due to the expectation that if those two bins were matched, any intermediate bin would also be matched. This reduced the numerical cost of performing the simulations.

It is our hope to extend future analyses to include more data. This would include data
from the RHIC beam-energy scan, from the LHC, and from the Cu+Cu, Cu+Au and U+U collisions at
RHIC. Data from the LHC is straight-forward to incorporate because the same theoretical
models can be used once one has added energy-dependence to the initial-state
parameterization. The Cu+Au and U+U data require significantly rethinking the
parameterization of the initial state, especially for Uranium due to large nuclear
deformation. Extending the analysis to include data from the beam-energy scan would
require significant changes to the model used here. At lower energies, one can no longer
assume Bjorken boost-invariance and can no longer ignore the baryon excess. Although our
present hydrodynamic code can work in three dimensions, significant theoretical work is
required to develop a parameterization for the three-dimensional initial state at
arbitrary beam energies.

\subsection{Initial Distillation of Observables}

Experimental collaborations have spent tremendous effort reducing the huge RHIC data set to a
finite number of published plots representing useful summaries of several classes of observables. For
example, the PHENIX and STAR collaborations have produced plots of proton spectra for
several centrality classes. Each plot might have a few dozen points. It is infeasible for
an emulator to reproduce each of these data points. Instead, each observable
was reduced to a few representative quantities. For a given species and centrality, spectra
were reduced to two numbers. The first number is the yield, or integrated spectra, within
a finite $p_t$ (transverse momentum) range. The ranges were set so that they ignored the
high $p_t$ tail, which is strongly affected by jets and is outside the scope of the
model. The second number would be the mean $p_t$ within that range. The choice to use mean
$p_t$ was motivated by a principal component analysis (PCA, described in next subsection) on the data points
within a spectra divided by the yield. This showed that 99\% of the variability of the spectra could be captured by two numbers.

\begin{figure}
\centerline{\includegraphics[width=0.48\textwidth]{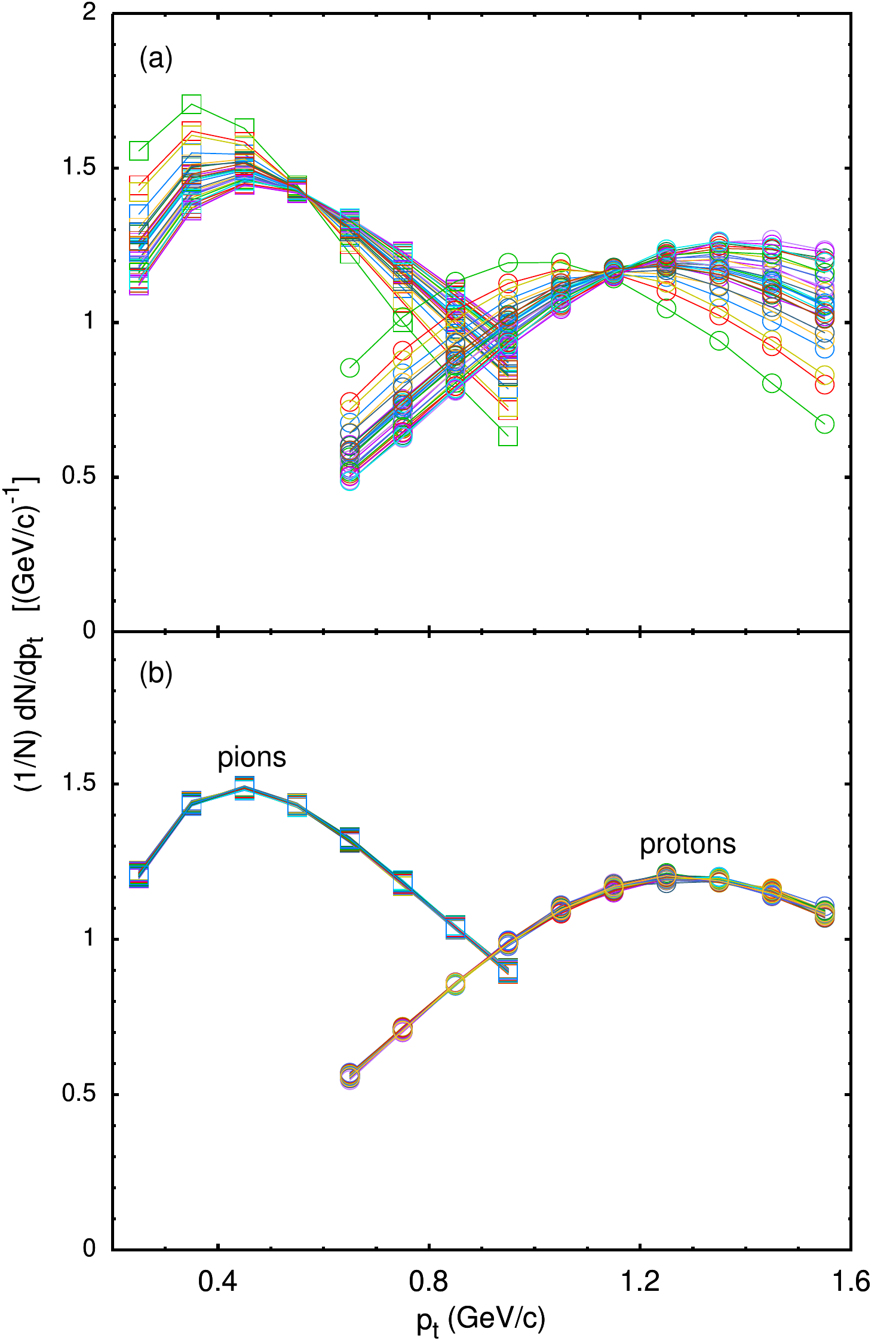}}
\caption{\label{fig:spectralshape}(color online) 
The probability density for creating either pions (squares) or proton (circles) of transverse momentum $p_t$ divided by the respective yields, i.e., spectral shapes, carries all the information in spectra outside what is described by yields. Spectral shapes for pions from 30 randomly chosen full model runs of the 729 performed runs used to sample the prior distribution are displayed in the upper panel. This demonstrates the variability of the spectral shapes throughout the parameter space. In the lower panel, 74 runs were chosen that had mean pion transverse momenta $573<p_t<575$ MeV/$c$. The fact that these calculations produce nearly indistinguishable spectral shapes shows that the mean transverse momenta encapsulates nearly all the variability in the spectral shapes over the prior parameter space. The same was done for proton spectra, with proton spectra from 30 randomly chosen model runs shown in the upper panel, and results from 44 runs whose mean proton transverse momentum was $1150<p_t<1152$ MeV/$c$ shown in the lower panel.}
\end{figure}

To illustrate the degree to which the yield and mean $p_t$ encapsulate the information carried by the spectra, calculations were selected from the initial 729 calculations that had the same mean transverse momentum within the acceptance window. In Fig. \ref{fig:spectralshape} the number of pions per unit transverse momentum are shown after being scaled by the net number of pions in the acceptance window, $(1/N)dN/dp_t$. With this scaling one can compare the shapes independent of the yields. In the upper panel, spectral shapes are shown for 30 randomly chosen calculations, while in the lower panel only those calculations with $573<p_t<575$ MeV/$c$ were used. These 74 runs should yield identical spectral shapes if $p_t$ carries the entire information carried by the spectral shapes. Also shown are 30 proton spectral shapes from random calculations in the upper panel, and 44 calculations where the mean transverse momentum of the protons was $1150<p_t<1152$ MeV/$c$. These calculations show that little, if any, additional information remains in the spectral shapes once one knows the mean $p_t$.

For the elliptic flow, the experimental information consists of plots of $v_2$ as a
function of $p_t$. A PCA analysis showed that the $p_t$-weighted value for $v_2$
effectively captured all the information within the set of model runs. The observable is defined by,
\begin{equation}
\langle\langle v_2\rangle\rangle = \frac{\sum_i v_{2,i}\langle p_{t}\rangle_i}{\sum_i \langle p_{t}\rangle_i}
\end{equation}
where the subscript $i$ refers to the transverse momentum bins in the STAR data, and $\langle p_{t}\rangle_i$ is the average $p_t$ of particles within that bin. This choice of binning reduces the degree to which two curves with the same $v_2$ vs. $p_t$ curves would differ if they had different spectra.

Femtoscopic information came from the STAR Collaboration, which analyzed the Gaussian radii ($R_{\rm
  out}$, $R_{\rm side}$ and $R_{\rm long}$) as a function of transverse
momentum. Simply averaging each radius over the several $p_t$ bins was found to
effectively encapsulate nearly all the variation of the femtoscopic radii throughout the
model runs.

In this manner the various experimental results were reduced to those listed in 
Table \ref{table:pcaobservables}. Each observable was also assigned an uncertainty. This
uncertainty represented the accuracy within which a comparison of the theoretically
determined value from a model run could be meaningfully compared to the corresponding
experimental measurement. Of all the observables in Table \ref{table:pcaobservables} only
$v_2$ has significant statistical error. The $v_2$ observable is also known to be
significantly affected by known shortcomings in the model, such as the lack of
event-by-event fluctuations. By averaging over many events with the same impact parameter, one can generate smooth initial conditions, which avoid the lumpy energy-density profiles caused by the finite number of colliding nucleons. The smooth conditions allow one to run only a single hydrodynamic evolution for the smoothed profile rather than running for many lumpy profiles. Finally, there are numerous schemes by which experimentalists determine $v_2$, which differ at the level of 5-10\%. In order to reduce non-flow correlations at the two- or three-body level, $v_2$ can be extracted from correlations of higher order \cite{Borghini:2001vi}. For non-identified particles, this has led to estimates of $v_2$ that are lower by approximately 10\% \cite{Adams:2004bi}. Since we are considering the $v_2$ of identified particles, and since the experimental four-particle-cummulant analysis has not been completed for identified particles, we compare our model to the two-particle result reduced by 10\%. Given the lack of fluctuations, it is rather difficult to choose which scheme one should compare with. For theses reasons, $v_2$ is assigned a larger percentage error than other observables for this study. For future analyses, especially those that include fluctuations, significant thought needs to be invested in determining a reasonable level of uncertainty for $v_2$.
\begin{table*}
\begin{tabular}{|c|c|c|c|c|c|}\hline
observable & $p_t$ weighting & centrality & collaboration & uncertainty & reduced uncertainty\\ \hline
$v_{2,\pi^+\pi^-}$ & ave. over 11 $p_t$ bins from 160 MeV/$c$ to 1 GeV/$c$ & 20-30\% & STAR$^*$ \cite{Adams:2004bi}& 12\% & 6\% \\
$R_{\rm out}$ & ave. over 4 $p_t$ bins from 150-500 MeV/$c$ & 0-5\% & STAR \cite{Abelev:2009tp}& 6\% & 3\%\\
$R_{\rm side}$ & ave. over 4 $p_t$ bins from 150-500 MeV/$c$& 0-5\% & STAR \cite{Abelev:2009tp}& 6\% & 3\% \\
$R_{\rm long}$ & ave. over 4 $p_t$ bins from 150-500 MeV/$c$& 0-5\% & STAR \cite{Abelev:2009tp}& 6\% & 3\% \\
$R_{\rm out}$ & ave. over 4 $p_t$ bins from 150-500 MeV/$c$& 20-30\% & STAR \cite{Abelev:2009tp}& 6\% & 3\% \\
$R_{\rm side}$ & ave. over 4 $p_t$ bins from 150-500 MeV/$c$& 20-30\% & STAR \cite{Abelev:2009tp}& 6\% & 3\% \\
$R_{\rm long}$ & ave. over 4 $p_t$ bins from 150-500 MeV/$c$& 20-30\% & STAR \cite{Abelev:2009tp}& 6\% & 3\% \\
$\langle p_t\rangle_{\pi^+\pi^-}$ & 200 MeV/$c < p_t< 1.0$ GeV/$c$ & 0-5\% & PHENIX \cite{Adler:2003cb}& 6\% & 3\% \\
$\langle p_t\rangle_{K^+K^-}$ & 400 MeV/$c < p_t< 1.3$ GeV/$c$ & 0-5\% & PHENIX \cite{Adler:2003cb}& 6\% & 3\% \\
$\langle p_t\rangle_{p\bar{p}}$ & 600 MeV/$c < p_t< 1.6$ GeV/$c$ & 0-5\% & PHENIX \cite{Adler:2003cb}& 6\% & 3\% \\
$\langle p_t\rangle_{\pi^+\pi^-}$ & 200 MeV/$c < p_t< 1.0$ GeV/$c$ & 20-30\% & PHENIX \cite{Adler:2003cb}& 6\% & 3\% \\
$\langle p_t\rangle_{K^+K^-}$ & 400 MeV/$c < p_t< 1.3$ GeV/$c$ & 20-30\% & PHENIX \cite{Adler:2003cb}& 6\% & 3\% \\
$\langle p_t\rangle_{p\bar{p}}$ & 600 MeV/$c < p_t< 1.6$ GeV/$c$ & 20-30\% & PHENIX \cite{Adler:2003cb}& 6\% & 3\% \\
$\pi^+\pi^-$ yield & 200 MeV/$c < p_t< 1.0$ GeV/$c$ & 0-5\% & PHENIX \cite{Adler:2003cb}& 6\% & 3\%\\
$\pi^+\pi^-$ yield & 200 MeV/$c < p_t< 1.0$ GeV/$c$ & 20-30\% & PHENIX \cite{Adler:2003cb}& 6\% & 3\%\\
\hline
\end{tabular}
\caption{\label{table:pcaobservables} Observables used to compare models to data. $^*$To account for non-flow correlations, the value of $v_2$ was reduced by 10\% from the value reported in \cite{Adams:2004bi}.}
\end{table*}

\subsection{Principal Component Analysis (PCA) of reduced observables}

One could create model emulators for each of the observables listed in
Table \ref{table:pcaobservables}. However, one can further distill the data to a handful of
principal components representing their most discriminating linear combinations. This serves
to further reduce the complexity of the emulator. Let $y_{\rm exp,i}$ and
$\sigma_i$ be data points and uncertainties for the $i=1$ through $N$ data points listed
in Table \ref{table:pcaobservables}. One then considers the corresponding quantities from
the model run $m$, $y_{m,i}$ where $m$ runs from 1 to the number of full model runs
$M$. A useful first step is to scale the quantities by their net uncertainty,
\begin{eqnarray}
\label{eq:ytilde}
\tilde{y}_{{\rm exp,i}}&=&\frac{y_{\rm exp,i}-\langle y_i\rangle}{\sigma_i},\\
\nonumber
\tilde{y}_{m,i}&=&\frac{y_{m,i}-\langle y_i\rangle}{\sigma_i},\\
\nonumber
\langle y_i\rangle&=&\frac{1}{M}\sum_{m=1}^M y_{m,i}.
\end{eqnarray}
The net uncertainties, $\sigma_i$ are operationally defined as the uncertainty involved in
comparing a model value to an experimental measurement. The measurements considered in
this paper are mainly limited by systematic uncertainties rather than those from finite statistics, and we assume that uncertainties are described by a normal distribution,
\begin{equation}
{\cal L}({\bf x})\sim \exp\left\{-\sum_i 
\frac{\left(y_i^{\rm(exp)}-y_i^{\rm(mod)}({\bf x})\right)^2}{2\sigma_i^2}
\right\},
\end{equation}
where $y^{\rm(exp)}$ and $y^{\rm(mod)}$ are the experimentally measured and model values
respectively. Even if the model parameters are exact, the models also have limited
accuracy due to shortcomings in the physics. Thus, the net uncertainty encapsulates both
theoretical and experimental uncertainties, i.e., they can be considered to describe the
inability of the model not only to describe the physics of the collision, but to also
account for the inadequacy of the model to describe uncertainties in the experimental
measurement and analysis.

The net uncertainties are listed in the last two columns of Fig. \ref{table:pcaobservables}. As described in the previous paragraph, systematic uncertainties for the models are insufficiently understood. For that reason, the calculation was repeated  with two choices for the uncertainty, a more pessimistic choice and a more optimistic choice with half the values. If only experimental uncertainties were considered, uncertainties would likely be stated at a few percent for most observables, and the more optimistic set of uncertainties would be more reasonable. The one exception would be the femtoscopic radii, where stated uncertainties are close to the pessimistic set. For instance, in STAR's femtoscopic data the outward and sideward data do not appear to approach one another at low $p_t$, where they differ by $\sim 5\%$. A more detailed analysis of experimental issues such as resolution might help clarify this issue.

Determining systematic uncertainties is usually difficult. For experimental systematic uncertainty, the accuracy of the apparatus and the analysis procedure define the uncertainty. In some cases, accuracy can be understood by comparing the apparatus measurement to a known reference signal, and if there is a random element to that signal, e.g., the electronic amplifications fluctuate in a known manner, one can confidently state the systematic uncertainty. For large experiments with complex analyses, the systematic uncertainty relies on expert judgment. For example, the efficiency of a particle physics detector can be estimated with detailed Monte Carlo simulations. However, the simulation relies on numerous approximate treatments of the detector and of the response of the detector to various input. Often the calibration procedures differ greatly from the environment in which the experiment is run. The reliance on expert judgment is never fully satisfactory, even after involving discussions of numerous collaborators. 

For complex simulations, theoretical systematic uncertainty is also unavoidable. Whether the problem involves simulations of heavy-ion simulations, or the cosmology of the early universe, the physics is always approximate at some level. As stated above, the net uncertainty is currently determined by the confidence with which one trusts  theoretical models to reproduce reality, even if given the most correct choice of parameters. In order to confidently assign these uncertainties, one must make an inventory of main missing physics ingredients and to assess the degree to which such shortcomings of the model might affect the result. For instance, smooth initial conditions are known to affect elliptic flow at the 5-10\% level. Other shortcomings are related to the choice of hadronization scheme (perfectly thermal), a crude final-state correction to pion spectra to account for Bose effects, the lack of baryon annihilation and regeneration, the lack of bulk viscosity, possible mean-field effects in the hadronic stage, uncertainty of the Eq. of state, and possible changes to the initial conditions. In most of these cases, assessing the associated uncertainties involves running an improved model with the effects. At this point, the improved model should replace the one used in this study, and if possible, uncertainties of the treatments should be parameterized and the parameters should be varied. As an example, the current model ignores the non-thermalized perturbative QCD component which affects spectra at high $p_t$. One could add a component to the spectra that scales as $1/p_t^4$ with a parameterized magnitude constrained by studies of spectra at high $p_t$. As another example, uncertainties in how hadronization proceeds could be incorporated by assigning fugacities based on quark numbers of various hadron species. One might easily increase the number of theoretical parameters from the half dozen considered here to one or two dozen. Since the range of these additional parameters is constrained by prior knowledge, the uncertainty to the spectra deriving from the uncertainty of this component can be represented by the width of the prior distribution of this parameter. This uncertainty could then be neglected when assigning a ``systematic theoretical'' uncertainty to the mean transverse momentum. This same approach can also be applied to some experimental uncertainties. For instance, the detector efficiency could be a parameter, which would then reduce the amount of systematic experimental uncertainty one would assign to the final measured yields. Such parameters are often referred to as ``nuisance'' parameters, though parameters considered to be nuisances by some scientists might be considered to be extremely interesting to others.

Once the modeling is better understood, one can go beyond the rather ad-hoc assignment of uncertainties considered here. At that point the experimental uncertainties should dominate. Accurately representing such uncertainties would require conversation with the experimental community. The principal goal of this paper is to investigate the feasibility of a large-scale statistical analysis on RHIC data, and significant improvement is needed in the modeling and in the assessment of uncertainties before the results can be considered robust and rigorous. Nonetheless, the analysis is an improvement in the state of the art, and by considering two sets of uncertainties one is able to assess the potential of the method, and understand the degree to which the various parameters are constrained, or might be constrained once uncertainties are better understood. 

To proceed with the principal componenta analysis, one first calculates the sample covariance of the model values amongst the $M$ model runs,
\begin{equation}
S_{ij}=\frac{1}{M}\sum_{m=1}^M
\tilde{y}_{m,i}\tilde{y}_{m,j}.
\end{equation}
The $N$ eigenvalues of $S$ are $\lambda_i$, and the normalized
eigenvectors are $\hat{\epsilon}_{i,j}$. One can then consider new
variables, $z_{m,i}$ which are linear combinations of the original
$\tilde{y}_{m,i}$ along the various directions defined by the
eigenvectors,
\begin{equation}
z_{m,i}=\sum_j \hat{\epsilon}_{i,j}\tilde{y}_{m,j}.
\end{equation}
With this procedure, the model values, $\tilde{y}_{m,i}$, are rotated
into a basis where the values $z_{m,i}$ have a diagonalized variance
over the model runs,
\begin{equation}
\frac{1}{M}\sum_{m=1}^M
z_{m,i}z_{m,j}=\lambda_i\delta_{ij}.
\end{equation}
The values $z_{m,i}$ are known as principal components. Since the values $\tilde{y}$ were
scaled by the uncertainties, the components $\tilde{y}_i$ have uncertainties of unity, and
after rotation the values $z_i$ also have uncertainties of unity. Since the variance of
$z$ within the model runs is diagonal, one can state that those components for which
$\lambda_i\ll1$ can be ignored because they do not assist in discriminating
parameters. Further, the discriminating power is often dominated by the first few
principal components, i.e., those with the largest $\lambda_i$. 

\begin{figure}[ht]
  \centering
	  \includegraphics[width=0.4\textwidth]{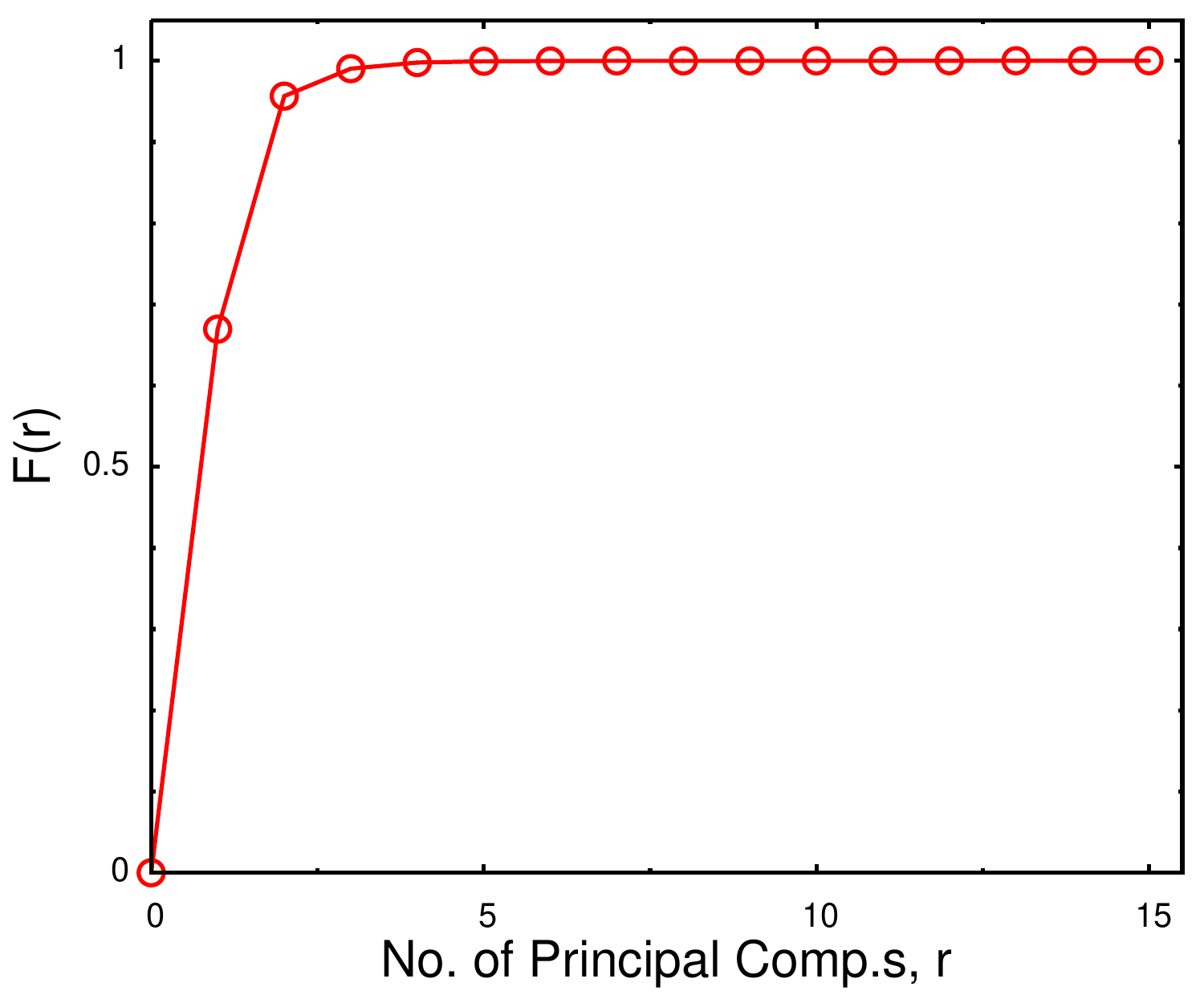}
  \caption{\label{fig:pca_resolving}(color online) The variance resolving power $F(r)$ of the principal
    components, only the first few components are needed to explain almost all of the
    observed variance.}
\end{figure}

To further justify our selection of principal components we show a plot of the normalized
cumulative variance explained by the largest $r$ components in Fig. \ref{fig:pca_resolving},
i.e
\begin{equation}
F(r) = \frac{\sum_{i=1}^{r} \lambda_i }{\sum_{i=1}^{N} \lambda_i},
\end{equation}
where we have sorted the eigenvalues into descending order.  Examination of this figure
clearly shows that the first three principal components are sufficient to explain almost
all of the variability of the model throughout the parameter space.

Once the principal components, $z_{i}$, have been
determined, one can invert the transformations to find $y_i$ in terms of the $z_i$. The
components which do not contribute strongly to the total variance can be set to zero and
the resulting $y_i$s will not be appreciably affected. In this particular case these are
the components with $\lambda \ll 1$.  Thus, the statistical analysis need only emulate
those components with $\lambda_i\gtrsim 1$.

Given the 15 observables outlined in Table \ref{table:pcaobservables}, one could construct an
emulator for each observable. However, a PCA analysis of the 15 intermediate observables
shows that not more than three of the principal components vary appreciably throughout
the model runs. For these three components, the corresponding fluctuations, $\langle \delta
z_i^2\rangle$ were of order unity or greater, while the remaining components fluctuated
significantly less than unity. Thus, instead of tuning 15 emulators, only six principal components
were considered (even though only three were truly needed). It is instructive to list the
values of $\lambda_i$ and the decomposition of the main components. This is shown in
Table \ref{table:pca}.
\begin{table*}
\begin{tabular}{|r|c|c|c|c|c|c|}\hline
observable $~~~\backslash$$~~~~\lambda_i$ & 18.36 & 7.87 & 0.93 & 0.21 & 0.04 & 0.012\\ \hline
       cent0to5\_PHENIX\_spectraPION\_YIELD&  0.43202 & 0.52170 & 0.21636 & 0.56290 & 0.06883 & 0.35417\\
      cent0to5\_PHENIX\_spectraPION\_MEANPT&  0.10117 & 0.02647 & 0.37032 &-0.08869 & 0.09235 &-0.24640\\
      cent0to5\_PHENIX\_spectraKAON\_MEANPT&  0.10770 & 0.03291 & 0.37755 &-0.07459 & 0.06328 &-0.26766\\
     cent0to5\_PHENIX\_spectraPPBAR\_MEANPT&  0.04925 & 0.02192 & 0.16751 &-0.02131 &-0.05466 &-0.19039\\
                  cent0to5\_STAR\_ROUT\_PION& -0.01942 & 0.06908 &-0.31734 &-0.02968 & 0.72626 & 0.12886\\
                 cent0to5\_STAR\_RSIDE\_PION&  0.09148 & 0.09321 & 0.07972 & 0.05565 & 0.11943 &-0.07137\\
                 cent0to5\_STAR\_RLONG\_PION&  0.08413 & 0.09520 &-0.13599 & 0.37546 & 0.08521 &-0.50343\\
     cent20to30\_PHENIX\_spectraPION\_YIELD&  0.43743 & 0.49869 &-0.32721 &-0.56043 &-0.26805 &-0.01286\\
    cent20to30\_PHENIX\_spectraPION\_MEANPT&  0.07549 & 0.03028 & 0.33981 &-0.23142 & 0.28313 &-0.06472\\
    cent20to30\_PHENIX\_spectraKAON\_MEANPT&  0.08266 & 0.03721 & 0.34043 &-0.23645 & 0.27941 &-0.06785\\
   cent20to30\_PHENIX\_spectraPPBAR\_MEANPT&  0.03791 & 0.02697 & 0.14297 &-0.11517 & 0.03747 &-0.09339\\
         cent20to30\_STAR\_V2\_PION\_PTWEIGHT& -0.74299 & 0.65843 & 0.08846 &-0.03607 &-0.01531 &-0.06192\\
                cent20to30\_STAR\_ROUT\_PION&  0.02955 & 0.03296 &-0.30420 &-0.06375 & 0.43249 &-0.09820\\
               cent20to30\_STAR\_RSIDE\_PION&  0.08368 & 0.09367 &-0.01379 &-0.21381 & 0.08021 &-0.06598\\
               cent20to30\_STAR\_RLONG\_PION&  0.08974 & 0.08905 &-0.24592 & 0.19088 &-0.07458 &-0.62873\\
               \hline
\end{tabular}
\caption{\label{table:pca}
  The first six principal components. Since the variables were initially scaled by their uncertainties, the eigenvalues, $\lambda_i$, describe the resolving power of the components. Only the first $\sim 4$ components are significant, i.e., $\lambda\gtrsim 1$. The table also provides the decomposition of the principal components in terms of the 15 observables.}
\end{table*}
The eigenvalues $\lambda_i$ represent the resolving power of the various principal
components.

To gain an understanding of the degree to which the parameter space is constrained by each
measurement of $z_i$, one can consider the simple case where observables depend linearly on the parameters and where the prior distribution of parameters, $x_\alpha$, is distributed with unit variance according to a Gaussian
distribution, $\langle x_\alpha x_\beta\rangle=\delta_{\alpha\beta}$. In that case, the gradient
of each principal component,
\begin{equation}
(\nabla z_j)_\alpha\equiv \frac{\partial z_j}{\partial x_\alpha},
\end{equation}
form a set of orthogonal vectors because the covariance $\langle
z_iz_j\rangle=\lambda_i\delta_{ij}$ is diagonal,
\begin{eqnarray}
\label{eq:lambdaslope}
\langle z_iz_j\rangle=(\nabla z_i)_\alpha(\nabla z_j)_\beta\langle x_\alpha x_\beta\rangle=
\nabla z_i\cdot\nabla z_j=\lambda_i\delta_{ij}.
\end{eqnarray}
Thus, if each component of $z$ depends linearly on $x$, each principal component
constrains a separate direction in parameter space. One can then understand the resolving
power by considering the simple case with one principal component and one parameter. Given
a measurement $z_{\rm(exp)}$ and assuming that the prior has unit variance and that $z$
depends linearly with $x$,
\begin{equation}
P(x)\sim e^{-x^2/2}\exp\left\{-
\left(mx-z_{\rm(exp)}\right)^2/2
\right\},
\end{equation}
where $dz/dx$ is the slope $m$, and from Eq. \ref{eq:lambdaslope} $m^2=\lambda$. Completing the squares in the argument of the exponential,
\begin{equation}
P(x)\sim e^{-(\lambda +1)(x-\mu^2)/2},
\end{equation}
where $\mu$ is the prior mean for $x$. This shows that if the response is purely linear, each principal component reduces the width of the posterior relative to the prior by a factor $1/\sqrt{1+\lambda_i}$. 

The first principal component in Table \ref{table:pca} carries the bulk of the resolving
power. Since $\lambda_1=18.36$, the linear considerations above suggest that a measurement
of the first principal component should constrain the original parameter space by a factor
of roughly $1/\sqrt{19.36}$. The second and third principal components also significantly
narrow the parameter space. All together, one expects these measures to constrain the fit
to on the order of 5\% of the original six-dimensional parameter space at the
``one-sigma'' level. This estimate of the resolving power is based on an assumption that
$z$ varies linearly with $x$, but nonetheless provides a useful, although crude,
expectation for how our analysis might ultimately constrain the parameter space.

The first and second components dominantly consist of measures of the multiplicity and of
the $v_2$ observable. This is not surprising. It shows that the most important aspect of
fitting data is to fit the multiplicity and elliptic flow. The third component has a large
mixture of $\langle p_t\rangle$ and interferometric observables. Thus, before performing
the parameter space exploration, one expects that those parameters driving the
multiplicity and elliptic flow will be the most significantly constrained.

\section{Theory of Model Emulators}
\label{sec:stat-theory}

Determining the posterior distribution of parameters can be stated within the context of Bayes theorem,
\begin{equation}
P(\bm{x}|\mathcal{O})=\frac{P(\mathcal{O}|\bm{x})P(\bm{x})}{P(\mathcal{O})}.
\end{equation}
Here, our goal is to determine the probability, $P(\bm{x}|\mathcal{O})$, of the parameters $\bm{x}$ being correct given the observations $\mathcal{O}$. The probability of the observations $\mathcal{O}$ being observed given the parameters $\bm{x}$ is $P(\mathcal{O}|\bm{x})$, and is determined by running the model with parameters $\bm{x}$ and comparing the model output to observations. If one assumes the uncertainties are of a Gaussian nature the conditional probability has a simple form,
\begin{equation}
P(\mathcal{O}|\bm{x})\sim \exp\left\{\sum_i\frac{(\mathcal{O}_{i,{\rm exp}}-\mathcal{O}_{i,{\rm mod}}(\bm{x}))^2}{2\sigma_i^2}\right\},
\end{equation}
where the experimental observation is $\mathcal{O}_{i,{\rm exp}}$, and the model prediction is $\mathcal{O}_{i,{\rm mod}}(\bm{x})$. Of course, one can choose different forms for $P(\mathcal{O}|\bm{x})$ depending on the circumstance. The Bayesian prior, $P(\bm{x})$, describes the probability of the parameter $\bm{x}$ in the absence of any information from the observables. Examples of a prior distribution might be a uniform distribution within a given range, or a normal distribution. The denominator, $P(\mathcal{O})$, is the probability of the the experimental measurement without having compared to the model, and given that the observation is known, can be treated as a constant. Markov-chain Monte Carlo procedures provide a list of points in parameter space weighted proportional to the likelihood, $\mathcal{L}(\bm{x})=P(\bm{x}|\mathcal{O})$, i.e. the posterior distribution. Since determining this distribution requires only the relative likelihoods of points, the denominator, $P(\mathcal{O})$, is irrelevant since it does not depend on $\bm{x}$. Further, for the calculations in this study we assume uniform priors, $P({\bm x})$ is a constant within a given range. With this choice $P({\mathcal O}|{\bm x})$ and $P({\bm x}|\mathcal{O})$ are effectively interchangeable.

The analysis here uses a Metropolis algorithm to produce the posterior distribution, and is a random walk in parameter space where each step is accepted or rejected according to the relative likelihood \cite{KooninCompPhys}. If the relative likelihood is higher, the step is accepted, whereas if it is lower the step is accepted with a probability of the relative likelihoods. According to the Ergodic theorem, this produces a ``time''-average of the distribution consistent with the likelihood. By ignoring the first section of the MCMC trace, referred to as the ``burn-in'', and by using a sufficiently large number of random steps, the sampling of points provides the means to not just determine the average of any parameter value as taken from the posterior, but can also find correlations between parameters, and should even identify likelihood distributions with multiple maxima. The method was tested by repeating with different starting points, and by visualizing the progress of the trace.

Developing an understanding of a six-dimensional parameter space requires hundreds of thousands of MCMC steps. Each step requires calculating the likelihood, which in turn requires running the full model. Running a complex code for each sampled point in parameter space is impractical. An alternative strategy has been to develop model emulators. Emulators effectively interpolate from an initial sampling of runs through the space. One may need hundreds or thousands of full model runs to tune, or train, an emulator. If one can afford to run the model for hundreds of times, and if the emulation is accurate, model emulators can be extremely effective. Models that have a smooth, or even monotonic, dependence on parameters are especially good candidates for emulation since fewer sampling points are required to provide a good base for interpolation. 

We construct a Gaussian Process emulator \cite{OHag:2006,Oakl:OHag:2002,Oakl:OHag:2004,Kenn:OHag:2000}, which acts as a statistical model of our computer model. An emulator is constructed by conditioning a prior Gaussian Process on a finite set of observations of model output, taken at points dispersed throughout the parameter space. Once the emulator is trained it can rapidly give predictions for both model outputs and an attendant measure of uncertainty about these outputs at any point in the parameter space.  This is a probability distribution for the model output at all points in parameter space and is by far the most useful feature of Gaussian Process emulators. The most common interpolation schemes, such as interpolating polynomials, produce an estimate of the model output at a given location in the parameter space with no indication as to the extent that this value should be trusted. Furthermore, numerical implementations of Gaussian Process emulators are computationally efficient (producing output in fractions of a second whereas the full model might require many minutes, hours, or days), making it feasible to predict vast numbers of model outputs in a short period of time. This ability opens new doors for the analysis of computer codes which would otherwise require unacceptable amounts of time \cite{Higd:Gatt:etal:2008,Baya:Berg:Paul:etal:2007}.

\begin{figure}
  \centering
	\includegraphics[width=0.5\textwidth]{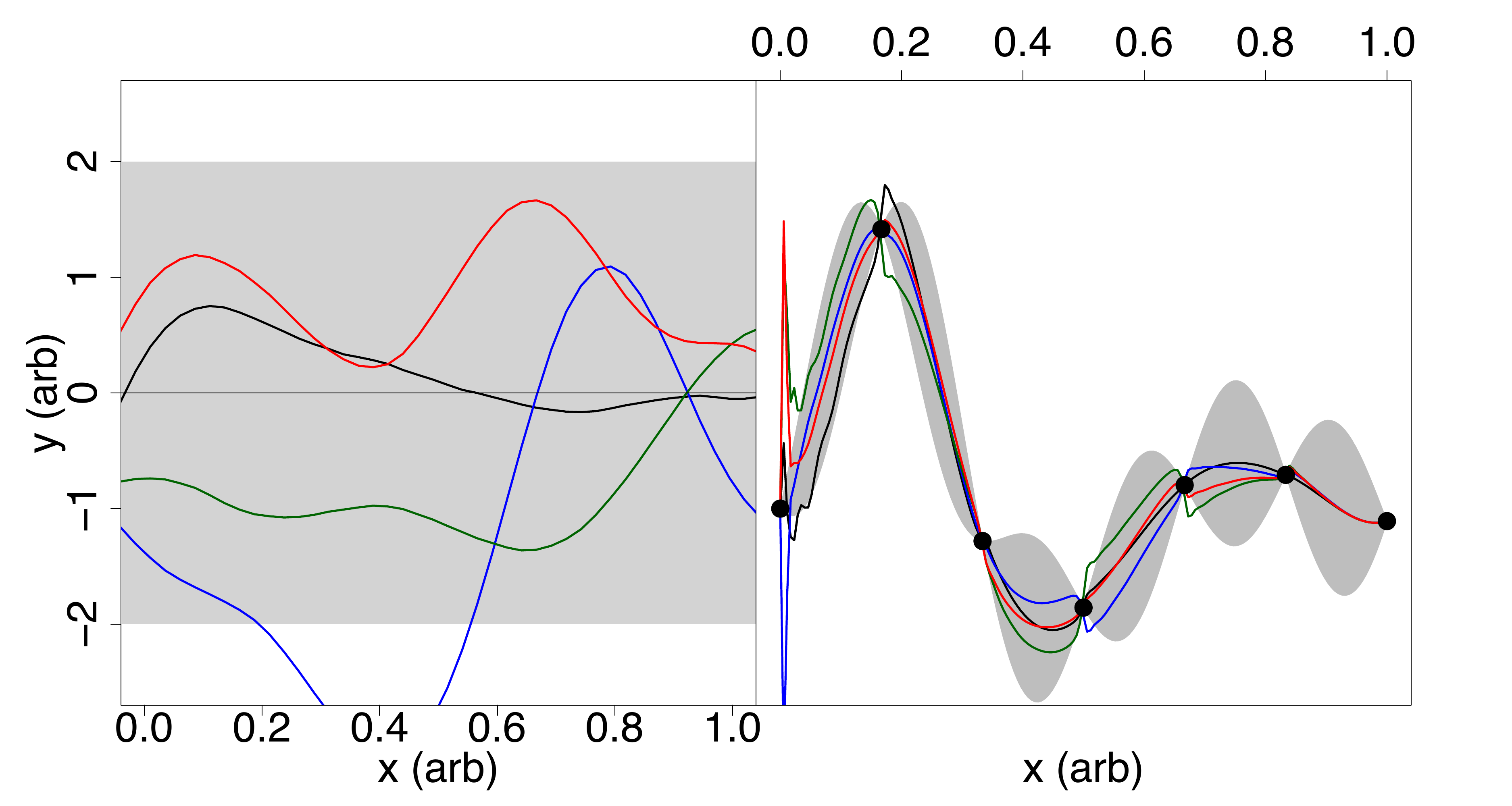}
  \caption{(color online) Left panel: Unconditioned draws from a Gaussian Process $GP(0, 1)$ with a mean
    of zero and constant unit variance. Right panel: draws from the same process after
    conditioning on 7 training points (black circles).  The gray band in both panels is a
    pointwise $95\%$ confidence interval.  Note how the uncertainty in the right panel
    grows when away from the training points. Refer to text for further details.}
\label{fig-gp-example}
\end{figure}

We construct an emulator for a model by conditioning a Gaussian Process prior (see
Fig. \ref{fig-gp-example}) on the training data \cite{Chil:Delf:1999,Cres:1993,Rasmussen}. A
Gaussian Process is a stochastic process with the property that any finite set of samples
drawn at different points of its domain will have a multivariate-normal (MVN)
distribution.  Samples drawn from a stochastic process will be functions indexed by a
continuous variable (such as a position or time) as opposed to a collection of values as
generated by, e.g., a normally-distributed random variable.  A Gaussian Process is
completely specified in terms of a mean and covariance, both of which can be functions of
the indexing variable $x$. The covariance, $c(x_1,x_2)$, might be any positive-definite
function of $x_1$ and $x_2$. An example of unconditioned draws is shown in the left panel
of Fig. \ref{fig-gp-example} for the case where the covariance depends only on $x_1-x_2$ and
is a power-exponential covariance function with unit length. The draws are smooth
functions over the domain space, and if enough samples are drawn from the process the
average of the resulting curves at each point would converge to zero.

A predictive distribution for the value of a computer model at new points in the design
space can be obtained by conditioning this process on a set of training points obtained
from running the model. Conditioning forces samples drawn from the process to always pass
through the training points. The resulting curves interpolate the training data, as shown
in the right hand panel of Fig. \ref{fig-gp-example}.  Repeated draws from the conditioned
posterior distribution would on average follow the underlying curve with some variation,
shown by the gray confidence regions.  These confidence bubbles grow away from the
training points, where the interpolation is least certain, and contract to zero at the
training points where the interpolation is absolutely certain. The posterior distribution
can be evaluated to give a mean and variance at any point in the parameter space.  We may
interpret the mean of the emulator as the predicted value at a point, the variance at this
point gives an indication of how close the mean can be expected to be to the true value of the model.

\begin{figure}
  \centering
	\includegraphics[width=0.45\textwidth, clip=true]{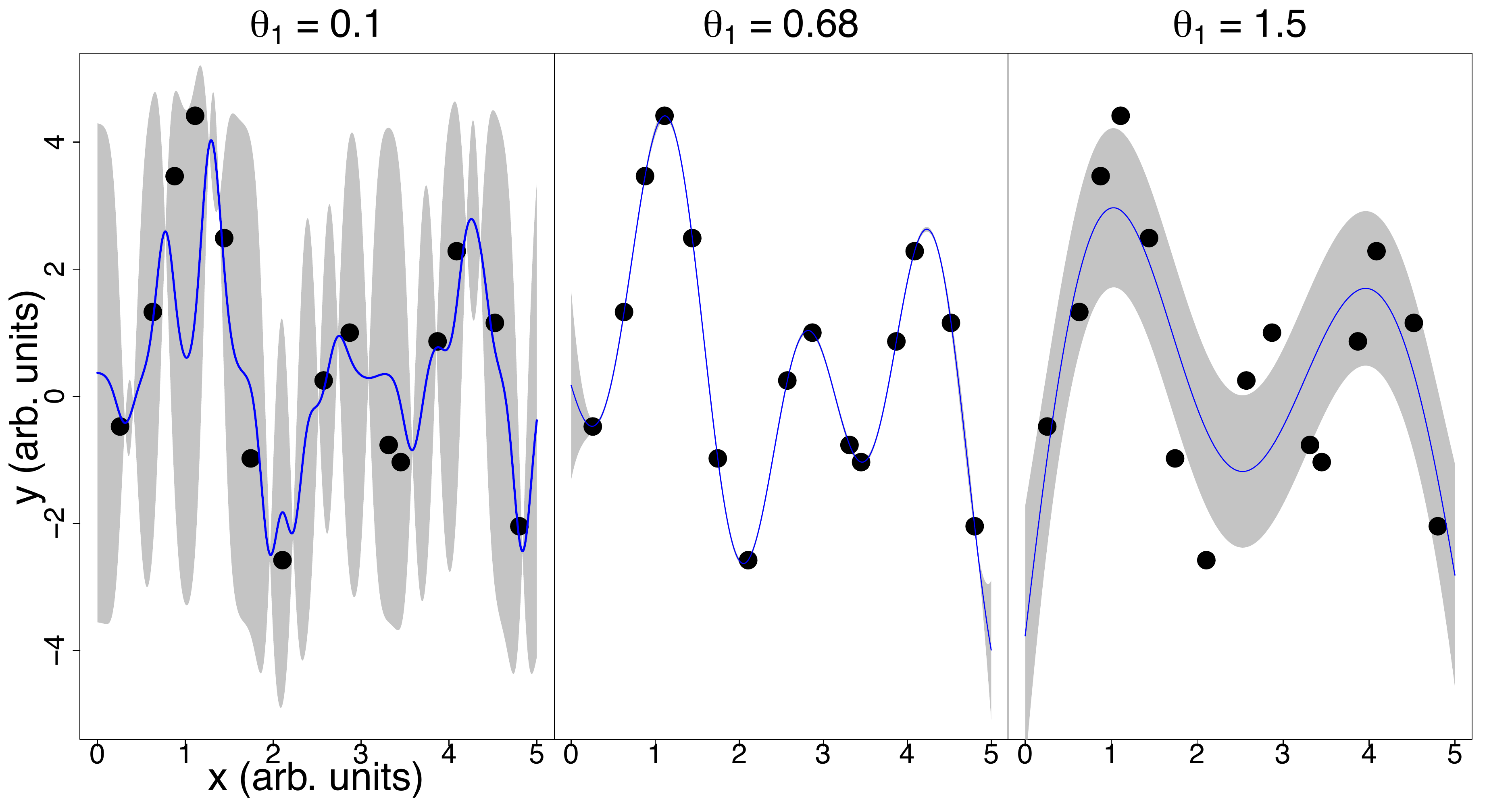}
  \caption{(color online) Demonstration of emulator behavior as a function of
    correlation length, $\theta_1$.  In all panels, the solid  blue  line shows  the  mean of  the
    emulator and the solid gray region is a $95\%$ confidence interval
    around  this region.  Left panel: fitting with a value of
    $\theta_1$ that is too
    small (under-smoothing).   Right panel: shows  over-smoothing by
    using a value of $\theta_1$ that is too  large.  Central panel:
    smoothing with a value  of  $\theta_1=0.68$  that was obtained  by  a maximum likelihood
    estimation method.}
  \label{fig-gp-theta}
\end{figure}

To construct an emulator we need to fully specify our Gaussian Process (GP) by choosing a
prior mean and a form for the covariance function. The model parameter space is taken to
be $p$-dimensional.  We model the prior mean by linear regression with some basis of
functions $\bm{h}(x)$. In this analysis we use the trivial basis $\bm{h}(x) =
\{ 1 \}$.  We specify a power exponential form for the covariance function with power
$\alpha \simeq 2$ to ensure smoothness of the GP draws ($\alpha$ has to be in $[1,2]$ to ensure positive definiteness),
\begin{widetext}
 \begin{equation}
   \label{eqn-emu-cov}
   c(\bm{x}_i, \bm{x}_j) = \theta_0 \exp\left(-
     \sum_{k=1}^{p} \left\{\frac{x_i^{k} -
       x_j^{k}}{\theta^{k}}\right\}^{\alpha}\right) + \delta_{ij} \theta_{N}, \quad \alpha\in[1,2].
\end{equation}
\end{widetext}
 Here, $\theta_0$ is the overall variance, the $\theta^{k}$ set characteristic length
 scales in each dimension in the parameter space and $\theta_N$ is a small term, usually
 called a nugget, added to ensure numerical convergence or to model some measurement error
 in the code output. The shape of the covariance function sets how the correlations
 between pairs of outputs vary as the distance between them in the parameter space
 increases.  The scales in the covariance function $\theta^{k}$ are estimated from the
 data using maximum likelihood methods \citep{Rasmussen}, in Fig. \ref{fig-gp-theta} we
 demonstrate their influence on an artificial data set. The linear regression model
 handles large scale trends of the model under study, and the Gaussian Process covariance
 structure captures the residual variations.

 Given a set of $n$ design points $\mathcal{D} = \{\bm{x}_1, \ldots, \bm{x}_n\}$ in a
 $p$-dimensional parameter space, and a set of $n$ training values representing the model
 output at the design locations $\bm{Y} = \{y_1, \ldots, y_n\}$ , the posterior
 distribution defining our emulator is
\begin{equation}
\mathcal{P}(\bm{x}, \bm{\theta}) \sim \mbox{GP}\left(\hat{m}(\bm{x}, \bm{\theta}), \hat{\Sigma}(\bm{x}, \bm{\theta})\right),
\end{equation}
for conditional mean $\hat{m}$ and covariance $\hat{\Sigma}$.
\begin{align}
  \label{eqn-emu-mean-var}
  \hat{m}(\bm{x}) &= \bm{h}(\bm{x})^{T}\hat{\beta} +
  \bm{k}^{T}(\bm{x}) \mathbf{C}^{-1} ( \bm{Y} - {\bf H} \hat{\beta}), \notag \\
  \hat{\Sigma}(\bm{x}_i, \bm{x}_j) &= c(\bm{x}_i, \bm{x}_j)  - \bm{k}^{T}(\bm{x}_i) \mathbf{C}^{-1} \bm{k}(\bm{x}_j) + \Gamma(x_i, x_j), \notag\\
  \mathbf{C}_{ij} &= c(\bm{x}_i, \bm{x}_j) \\
  \Gamma(x_i,   x_j)  &=  \left(   \bm{h}(\bm{x_i})^{T}  -
    \bm{k}^{T}(\bm{x_i})\mathbf{C}^{-1}               {\bf
      H}\right)^{T}
  \left({\bf H}^{T} \mathbf{C}^{-1} {\bf H}\right)^{-1} \notag \\
  &\left(\bm{h}(\bm{x_j})^{T} -
    \bm{k}^{T}(\bm{x_j})\mathbf{C}^{-1}{\bf H} \right), \notag \\
  \bm{k}(\bm{x})^{T}  &= \left(  c(\bm{x}_1, \bm{x})  ,  \ldots, c(\bm{x}_n,
    \bm{x}) \right).
\end{align}
Where     $\hat{m}(\boldmath{\it x})$    is     the     posterior    mean at $\bm{x}$,
$\hat{\Sigma}(\bm{x}_i,   \bm{x}_j)$   is  the   posterior
covariance between points $\bm{x}_i$ and $\bm{x}_j$, $\mathbf{C}$ is the $n \times n$ covariance matrix of the design $\mathcal{D}$,  $\hat{\beta}$  are the  maximum-likelihood  estimated
regression  coefficients,  $\bm{h}$   the  basis  of  regression
functions and ${\bf H}$ the matrix of these functions evaluated at the
training points.  

The elements of the vector $\bm{k}(\bm{x})$ are the covariance of an output at $\bm{x}$
and each element of the training set.  It is through this vector $\bm{k}(\bm{x})$ that
the emulator ``feels out'' how correlated an output at $\bm{x}$ is with the training set and
thus how similar the emulated mean should be to the training values at those points.  Note
that the quantities defined in Eq. \ref{eqn-emu-mean-var} depend implicitly upon the choice
of correlation length scales $\bm{\theta} = \{ \theta_0, \theta^{k}, \theta_N\}$
which determine the shape of the covariance function.

The expression for the mean in Eq. \ref{eqn-emu-mean-var} can be decomposed into a
contribution from the prior, the linear regression model
$\bm{h}(\bm{x})^{T}\hat{\beta}$ plus a contribution applied to the residuals
determined by the covariance structure $\bm{k}^{T}(\bm{x})\mathbf{C}^{-1}(
\bm{Y} - {\bf H} \hat{\beta})$.  Similarly the covariance can be decomposed into a
contribution from the prior, the covariance function $c(\bm{x}_i, \bm{x}_j)$
plus corrections arising from the prior covariance structure and the covariance of the new
location $\bm{x}$ through $\bm{k}(\bm{x})$. These terms weight the points $\bm{x}_i,
\bm{x}_j$ more highly the closer they are to the training points through
$\bm{k}$.  The $\Gamma$ term gives the corrections to the covariance arising from
the regression model.

In our study, we run the full code at $N=729$ (chosen because $3^6=729$) points from the
parameter space. A Latin Hyper Cube (LHC) design is used to generate the training
locations in the parameter space. This is an efficient design for space-filling in high
dimensional parameter spaces. This is an efficient sparse design for high dimensional parameter spaces that is ``space-filling'' in the sense that all its lower-dimensional projections are distributed as evenly as possible  \cite{Sack:Welc:Mitc:Wynn:1989,Sant:Will:Notz:2003,McKay}.


The output from the model code is multivariate. Although fully multivariate emulator
formulations do exist they are challenging to implement. Instead we follow the now
somewhat standard procedure of creating emulators for some decomposition of the code
output see eg \cite{Baya:Berg:Paul:etal:2007a, Higd:Gatt:etal:2008}. In this case we apply
a principle components decomposition to the model output and build emulators for each
significant component as detailed above. 


\section{Testing the Emulator}

The goal of this section is to investigate the reliability and accuracy of the Gaussian-process emulator described in the previous section. Tuning the Gaussian-process emulator involves choosing the hyper-parameters described in Eq. \ref{eqn-emu-cov}. The success of the tuning was determined by comparing emulated data to model predictions from 32 runs performed at points in parameter space not used to tune the model. These 32 points in parameter space were chosen randomly from the six-dimensional space.

The first attempt at finding optimized hyper-parameters used the same methods of \cite{Gomez:2012ak,Rasmussen}. However, that approach was not robust, and often led to inaccurate emulators. A more accurate result ensued by simply setting the hyper-radii, the $\theta^i$ values in Eq. \ref{eqn-emu-cov}, equal to half the range for each parameter $x^i$ in the model space. The exponent $\alpha$ was set to 1.5 and the nugget $\Theta_0$ was set to zero. Changing the hyper-radii by factors of two, or adjusting the exponent anywhere between 1.0 and 2.0 had little effect. For perspective, competing interpolating schemes were constructed, one based on a quadratic fit, and a second based on a linear fit where neighboring points were more heavily weighted in the fits. Each of these schemes was slightly less accurate than the Gaussian process emulator with the hyper-parameters chosen as described above. However, all these procedures performed better than the Gaussian process emulator using Maximum-Likelihood-Estimation (MLE) hyper-parameters as described in \cite{Gomez:2012ak,Rasmussen}. This failure to find good hyper-parameters may come from the numerical challenges of the MLE optimization process given the large number of training points.

\begin{figure}
\centerline{\includegraphics[width=0.4\textwidth]{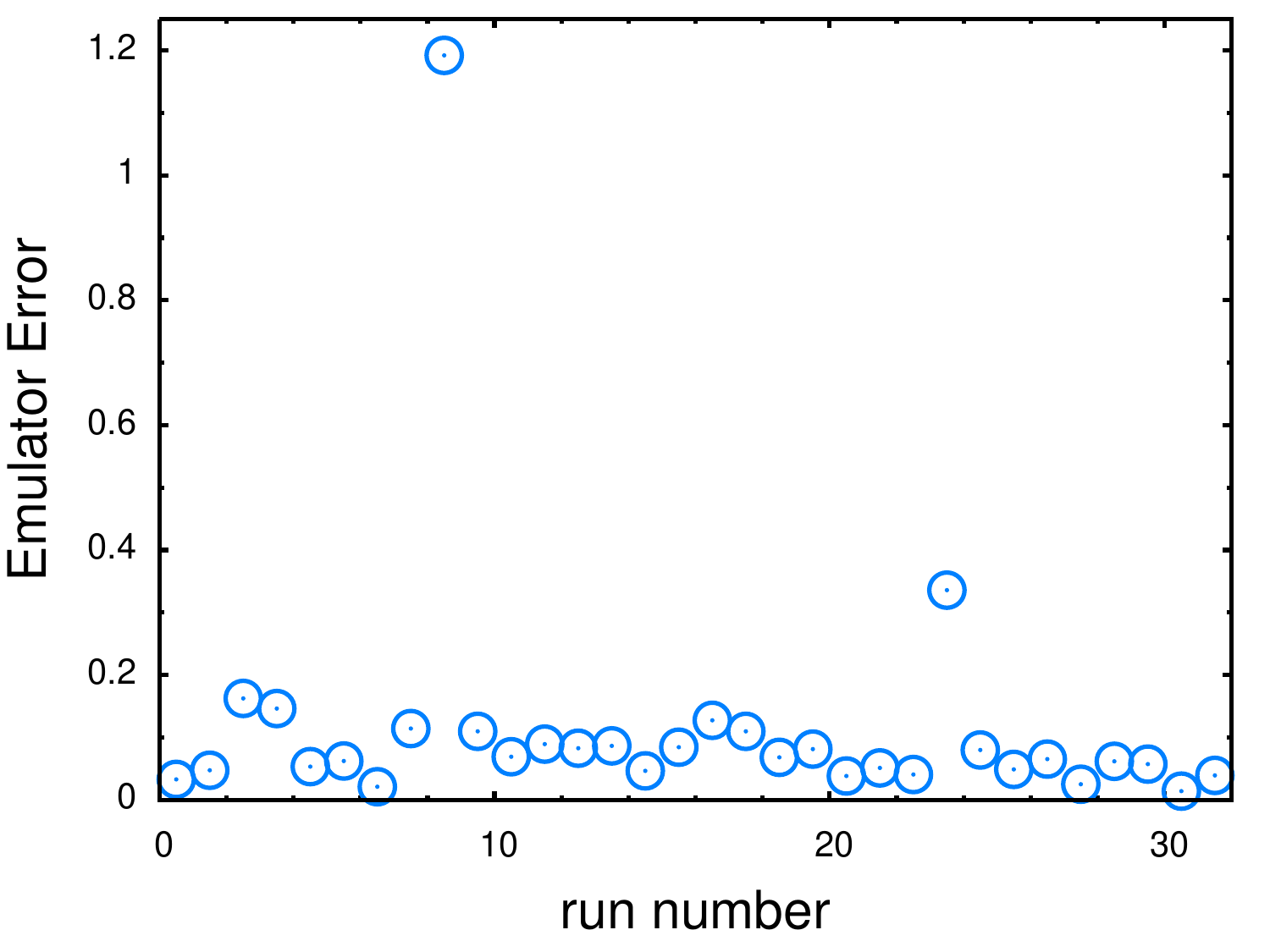}}
\caption{\label{fig:gpcheck}(color online) The emulator error, $\mathrm{EE}$, is shown for the 32
  test runs. If the emulator error per principal component were of order of the
  experimental and model uncertainties, the values of $\mathrm{EE}$ would be near six. The errors
  above are significantly smaller.}
\end{figure}

The Gaussian process emulator explicitly reproduces $z_i({\bf x})$ whenever ${\bf x}$
approaches one of the training points, ${\bf x}_n, n=1\ldots 729$. To test the
emulator points had to be chosen away from the training points, and 32 additional full
model runs were performed at random points throughout the parameter space. The emulator
error can be summarized as
\begin{equation}
\mathrm{EE}({\bf x})\equiv \sum_{i=1}^{r}\left(z^{\rm(emu)}_i({\bf x})-z_i^{\rm{\rm(mod)}}({\bf x})\right)^2,
\end{equation}
where $z^{\rm(emu)}_i({\bf x})$ is the conditional mean from the $i$'th emulator as given
by Eq. \ref{eqn-emu-mean-var} and $r$ is the total number of observation principle
components retained. A plot of this for the withheld data points is displayed in
Fig. \ref{fig:gpcheck} for the 32 test runs. By construction, $\mathrm{EE}({\mathbf x_n})$ is zero for the
729 training runs. The fact that the net errors were less than unity, even after summing
over six principal components, shows that the emulator did an outstanding job of
reproducing the data. Furthermore, the emulator error is of the order of the statistical
error of the model (which mainly comes from the calculation of $v_2$, which suggests that
in this case further improving the emulator would not significantly improve the final
result.

The Gaussian Process emulator was remarkably accurate, with the net error summed over all principal components being of order unity for only one of the 32 points, and much better for the other points. Two other emulator schemes were also investigated. The first was a quadratic fit weighted by the likelihood for each point. A second one was a linear fit, with the fit parameters chosen depending on the location in parameter space, and weighted more heavily with nearby points. All three of these methods provided accuracies similar to depicted in Fig. \ref{fig:gpcheck}, and all three led to nearly identical posterior distributions. The success of these emulators over a wide range of schemes and parameters is probably due to the smooth and monotonic response of the model to parameters. At high center of mass energies, the physical system is highly explosive. Within the range of
parameters considered the explosiveness is modified, but the behavior never changes qualitatively, and one expects a monotonic response to the parameters. This might not be as true at lower energies.

It was seen that the estimate of the errors of the emulation as defined in
Eq. \ref{eqn-emu-mean-var} often significantly underestimated the accuracy of the emulator
as tested in Fig. \ref{fig:gpcheck}. The net error tended to be less than a half unit, even though it was summed over multiple degrees of freedom. Since the error associated with the accuracy of the emulator was so small, the emulator error was incorporated into the calculation of the likelihood in a simplified manner. The uncertainty inherent to the data and models for a specific principal component was unity due to the choice in how to scale the
$z_i$ values. By adding in the emulator error, the total uncertainty should be
$\sigma^2=1+\sigma_e^2$ for each component. The likelihood used by the MCMC is then,
\begin{equation}
{\mathcal L}({\bf x})\propto \exp \left\{-\frac{1}{2}
\sum_i\frac{(z_i^{\rm(emu)}({\bf x})-z_i^{\rm(exp)})^2}{1+\sigma_e^2}\right\}.
\end{equation}
For our MCMC calculations, $\sigma_e$ was set to 0.1 according to an estimate of the error per degree of freedom from Fig. \ref{fig:gpcheck}. This increased the width of the posterior region of parameter space by only a few percent.

\section{MCMC Results}

As shown in the previous section, the emulator accurately reproduces the
log-likelihood. For the MCMC search the Gaussian process emulator was run sampling many
millions of points in parameter space. The trace provides an ergodic sample of the allowed
regions in parameter space, i.e., the posterior distribution. The MCMC procedure applied
here is a Metropolis algorithm. First, the parameter space was scaled and translated so
that it was centered around zero, and that the flat prior had unit variance, i.e., it
varied from $-\sqrt{3}$ to $+\sqrt{3}$.  First, a random point was chosen in the
six-dimensional parameter space ${\bf x}_1$, from which one takes a random step to ${\bf
  x}_2={\bf x}_1+\delta{\bf x}$. The random steps $\delta{\bf x}$ were chosen according to
a six-dimensional Gaussian with the step size in each dimension being 0.1. The likelihoods
were calculated for each point. If the likelihood ${\mathcal L}({\bf x}_2)$ was higher
than ${\mathcal L}({\bf x}_1)$, the step was accepted, and if the likelihood was smaller,
the step was accepted with the probability of the ratios of the two likelihoods. After the
100,000-step burn-in phase, the trace was stored by writing down every tenth point. The
resulting distribution is proportional to the likelihood \cite{KooninCompPhys} and
represents an ergodic sampling of the posterior distribution for a uniform prior. The
trace finished when $10^6$ points were written to disk. The procedure was repeated several times from different starting points to ensure the robustness of the trace. Visualization of the trace also appeared to show that the length of the search was sufficient. The ease with which the MCMC mapped out the posterior is probably explained by the lack of complex topology of the posterior distribution, i.e., we never found multiple maxima in the likelihoods as the dependence of the principal components with respect to the parameters appeared monotonic.

To evaluate the success of the emulation, 20 points were randomly chosen from the MCMC
trace and were then evaluated with the full model. The observables used for the original
analysis were then plotted for each of the 20 points in parameter space. Another twenty
points were chosen randomly from the original parameter space, i.e. they are consistent
with the flat prior distribution. Again, the observables were calculated with the full
model for each of these points in parameter space. One expects the observables for each of
the 20 points representing the MCMC trace to reasonably well match the experimental data,
while the points chosen randomly from the prior distribution should lead to a wider range
of observables, some of which should be inconsistent with the data.

\begin{figure*}
\centerline{\includegraphics[width=0.7\textwidth]{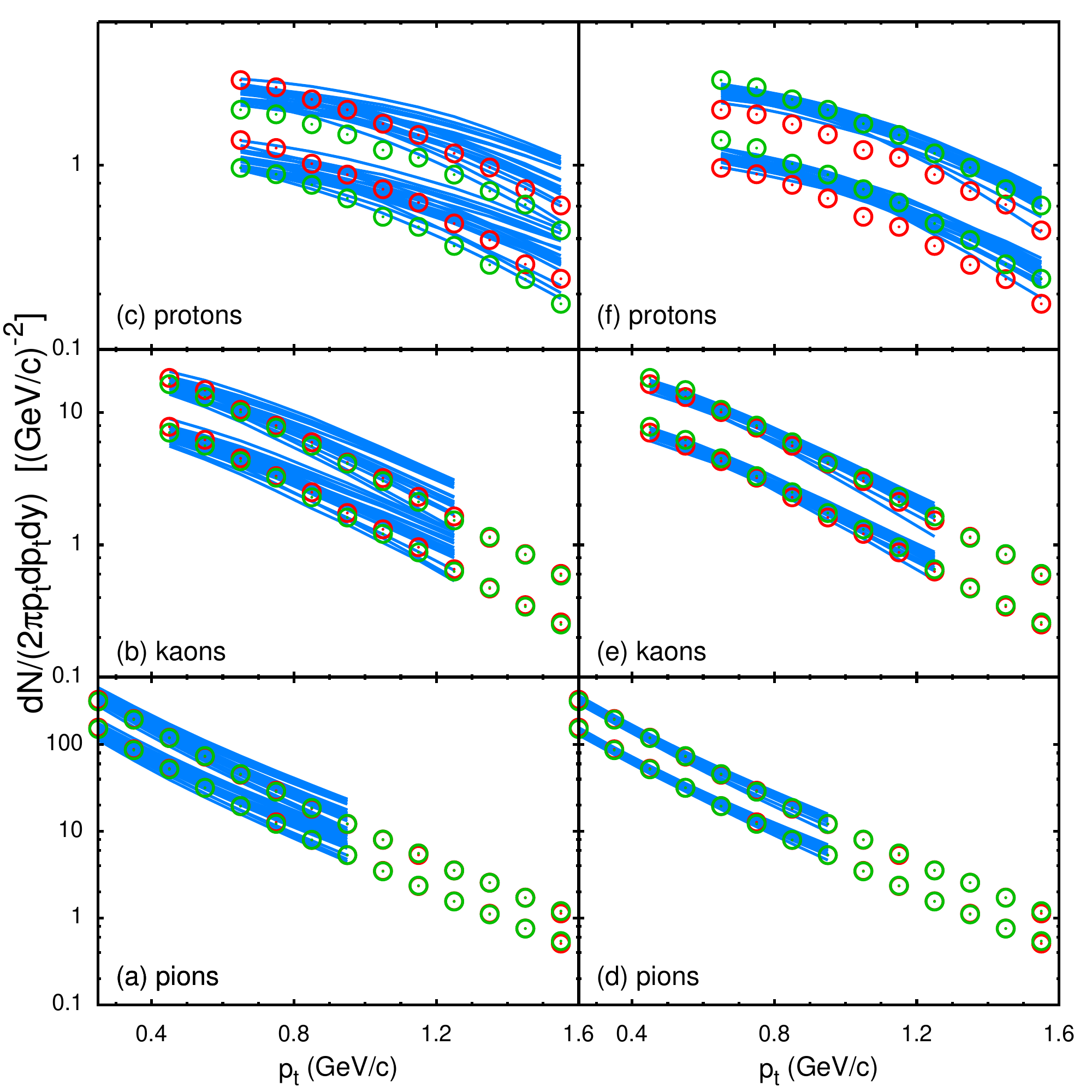}}
\caption{\label{fig:spectramcmc}(color online) Left-side panels: Pion, kaon and proton
  spectra from 20 model calculations where parameters are randomly chosen from the prior
  distribution. Model calculations are blue lines, and experimental data from PHENIX is
  shown as red/green circles for positive/negative charges. Results are shown for both 5\%
  most central, and for the 20-30\% centrality bin. Due to lack of some chemical
  reactions, normalizations for kaons and pions in the model were scaled by factors of
  0.85 and 0.6 respectively. Right panels: Same as left-side but with 20 model
  calculations where parameters were chosen randomly from posterior distribution as
  sampled by MCMC trace. }
\end{figure*}
Comparisons of the spectra from the model runs characterizing the prior and posterior
distributions are shown in Fig. \ref{fig:spectramcmc}. Parameters from the posterior
distributions lead to far superior fits, for both the yields and for the shape of the spectra. From
the figure, one can see that the spectra for heavier particles provide more discriminating
power. This comes from the greater sensitivity to collective flow, and emphasizes the
importance of having reliable measurements of proton spectra. At RHIC, STAR's proton
spectra \cite{Abelev:2008ab} are warmer than those of PHENIX \cite{Adler:2003cb}, and their estimate of the mean $p_t$ for protons
is 7\% higher. Whereas PHENIX shows the mean $p_t$ of protons staying steady or perhaps
slightly falling with increasing centrality, STAR's analysis show a rising mean $p_t$. If
the mean $p_t$ were indeed higher than what PHENIX reports, the extracted parameters
should change, e.g., the initial collective flow might come out higher.

\begin{figure}
\centerline{\includegraphics[width=0.42\textwidth]{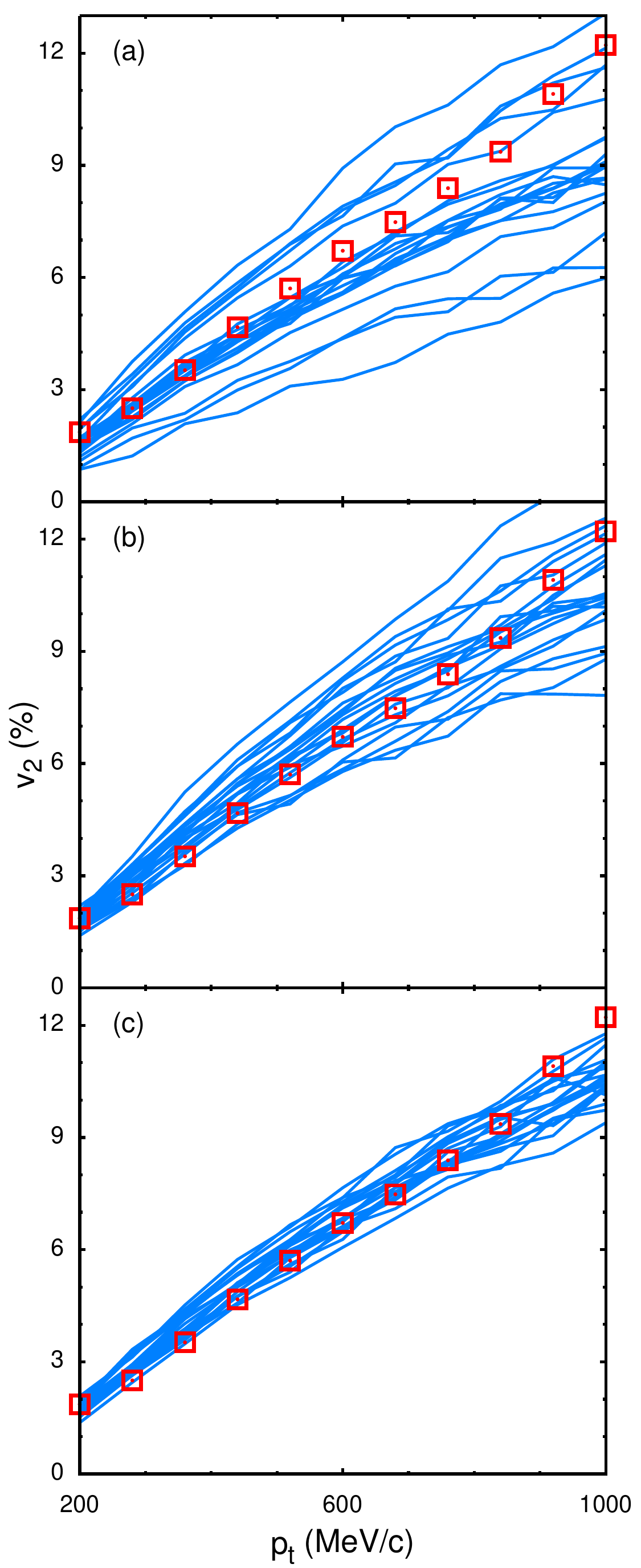}}
\caption{\label{fig:v2}(color online)
Upper panel (a): For 20 points in parameter space randomly chosen from the prior distribution, $v_2$ for pions is plotted as a function of $p_t$ for full model runs. Blue lines represent model calculations whereas are squares are experimental data. Middle panel (b): Same as upper panel, except 20 points are randomly taken from posterior distribution as sampled by the MCMC trace, and the lower-panel (c) shows results taken from calculations using parameters chosen from the posterior where the calculations had uncertainties reduced by a factor of 2.}
\end{figure}

Figure \ref{fig:v2} shows $v_2$ as a function of $p_t$ for identified pions as calculated from the
same representative points in parameter space for both the prior and posterior
distributions as were used for the spectra. The MCMC is clearly successful in identifying points in parameter space
that when run through the full model matched the experimental measurement of $v_2$. Further, given
that the systematic uncertainty of specifying the $p_t$ averaged $v_2$ was assumed to be
12\%, the spread of $v_2$ vs. $p_t$ plots appears consistent with expectations.

Although the overall trend of the source radii were matched by the model, a consistent discrepancy
between the data and model calculations using parameters from the posterior distribution
is evident. At low $p_t$, the sideward source sizes is over-predicted by
approximately 10\%, which is about double the expected systematic error.  The longitudinal source sizes are consistently over-predicted by the model. A $\sim 5\%$ overprediction was expected given the lack of longitudinal acceleration inherent to the assumption of boost-invariance used in the calculations \cite{Vredevoogd:2012ui}. Additionally, the finite longitudinal size might also lead to an additional few percent decrease in the longitudinal radii. Other aspects of the approximation, such as in how the $\pi\pi$ Coulomb interaction was treated, or in the approximation of independent emission used in the Koonin formula may have affected the answer at the level of a few percent. Finally, the procedure of extracting Gaussian radii from correlation functions can affect the answer. Since the actual correlations are not Gaussian, the fitted radii can depend on how various parts of the correlation function are weighted in the fit \cite{Frodermann:2006sp}. The calculations could be improved by using the same binnings and cuts as was used for the data, e.g., correlations at very small momenta are cut off experimentally due to two-track resolution issues.

\begin{figure}
\includegraphics[width=0.5\textwidth]{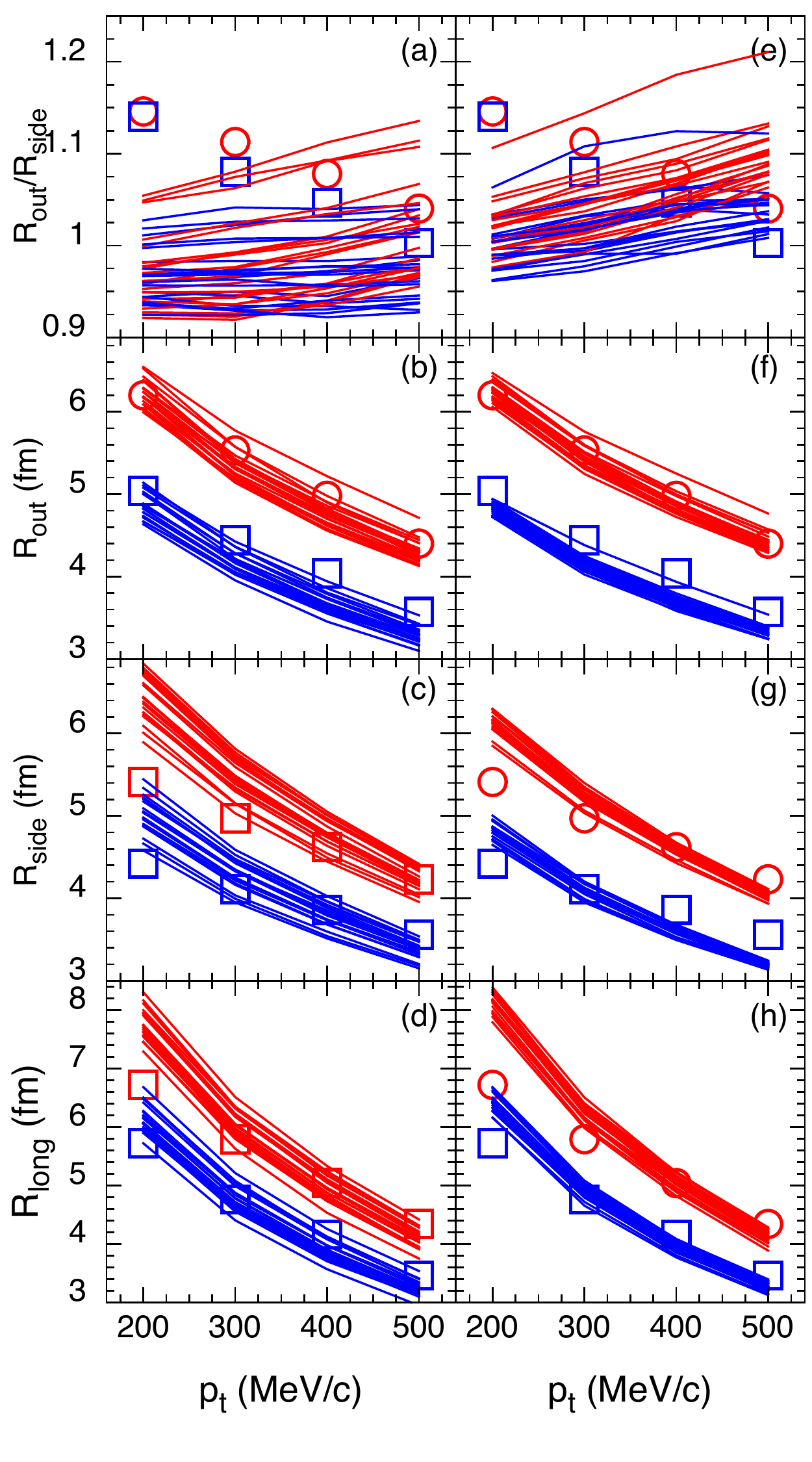}
\caption{\label{fig:hbt}(color online)
Femtoscopic radii are shown for calculations from the prior distribution (left half a-d) and from the posterior (right half, e-h) calculations. Red circles and blue squares refer experimentally extracted source radii from (0-5\%) and (20-30\%) centrality respectively. The red and blue lines show the corresponding theoretical calculations. The posterior calculations well reproduce the sideward and outward radii except at low $p_t$. The longitudinal radii from the calculations are consistently larger than the experimental ones.}
\end{figure}

From analyticity, one expects that the $R_{\rm out}$ and $R_{\rm side}$ sizes should approach one
another as $p_t\rightarrow 0$. As can be seen in the upper panel of Fig. \ref{fig:hbt},
this does not appear to be holding true in the data. Either the lower range of $p_t$ (200
MeV/$c$) is not sufficiently small, or an acceptance/efficiency effect in the detector is
affecting the result. This issue should be resolved if femtoscopic analyses are to be
applied with confidence near a 5\% level. However, since the femtoscopic observables carry a relatively small weight of the strongest principal components, as seen in Table \ref{table:pca}, resolving this puzzle is not expected to significantly change the extracted model parameters. From past experience, source radii are known to be sensitive to the equation of state, and for studies that very the equation of state, one expects femtoscopic observables to play a more critical role.

\begin{figure*}[p]
\centerline{\includegraphics[width=\textwidth]{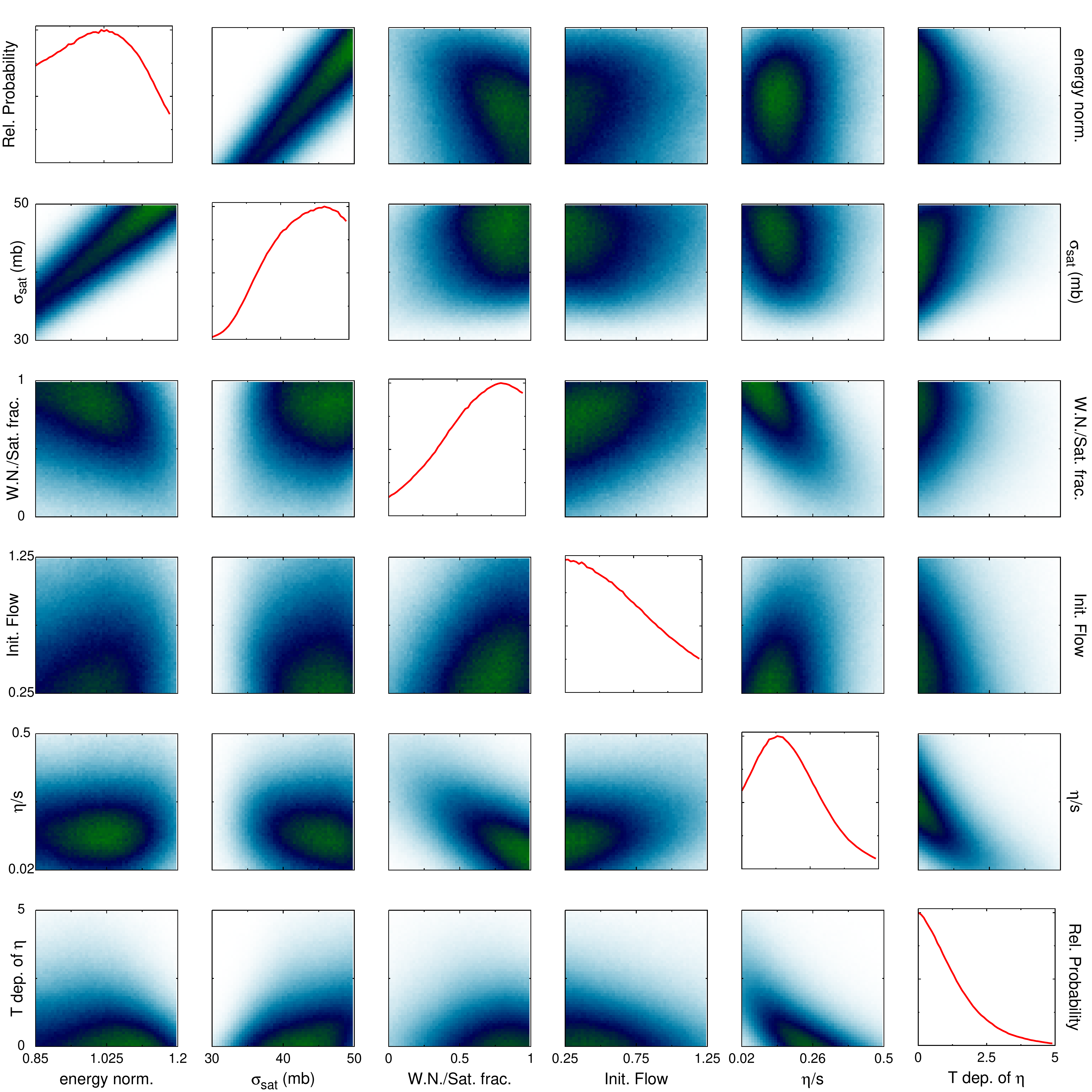}}
\caption{\label{fig:gpmcmc}(color online) The distribution of acceptable values for each
  of the six model parameters are shown along the diagonal. The off-diagonal plots display
  the correlation between all pairs of observables. Four of the six parameters refer to
  the initial state (hence the ``I.C.'' in their name) and the last two describe the shear
  viscosity. This calculation was based on the more pessimistic assumption of uncertainties in Table \ref{table:pcaobservables}. }
\end{figure*}

From the MCMC traces, the distribution of the various parameters and the correlations
between pairs of parameters are shown for the Gaussian-process emulator in
Fig. \ref{fig:gpmcmc}. The plots along the diagonal display the range of acceptable values
for individual parameters, integrated over all values of the other five
parameters. Although over 90\% of the six-dimensional parameter space is eliminated at the
one-sigma level, the individual parameters are rarely constrained to less than half their
initial range when other parameters are allowed to vary.

The first four parameters (``energy norm'', ``$\sigma_{\rm sat}$'', ``W.N./Sat. frac.'' and
``Init. Flow'') define the initial state of the hydrodynamic treatment. The first parameter
``energy norm'' sets the constant of proportionality between the product of the areal
densities of the incoming nuclei, and the initial energy density used to initiate the hydrodynamic treatment. In
the limit of low aerial densities this should be consistent with $pp$ collisions. Thus,
the range of the prior distribution was quite small, and the statistical analysis did
little to further constrain it. The parameter ``$\sigma_{\rm sat}$'' is defined in Eq. \ref{eq:wn} and parameterizes the saturation of the energy density with
multiple collisions. The preferred value appears rather close to the value of 42 mb
typically used in the wounded nucleon model, though there is a fairly wide range of
accepted values. The parameter ``W.N./Sat. frac.'' refers to $f_{\rm w.n.}$ in Eq. (\ref{eq:icweights}) and sets the weights between the wounded nucleon and the saturation parameterizations. This shows a
preference for the wounded nucleon prescription which gives a smaller initial anisotropy
than the saturation parameterization. The final initial-condition parameterization,
``Init. Flow'' sets the initial transverse flow set in the hydrodynamic calculation. The
parameter sets the initial flow as a fraction of the amount described by
Eq. \ref{eq:univflow}, which should be expected in the limit of high-energy. The MCMC trace
points to a rather small fraction of this flow, though like all of the initial-condition
parameters, the range of possible values is fairly broad.

The last two parameters define the viscosity. The viscosity at $T=170$ MeV is referred
to as ``$\eta/s$'' in Fig. \ref{fig:gpmcmc}, and the temperature dependence is labeled by
``$T$ dep. of $\eta$'', and refers to the parameter $\alpha$ in Eq. \ref{eq:alphadef}. Both
are significantly constrained as a fraction of the original parameter space. The range of
$\eta/s$ is consistent with similar, but less complete, searches through parameter space
using similar models \cite{Soltz:2012rk,Heinz:2011kt}. In \cite{Niemi:2011ix}, the authors
found little sensitivity to the viscosity at higher temperatures, but considered a smaller
variation of the viscosity with temperature than was considered here.

Figure \ref{fig:gpmcmc} also displays cross-correlations from the MCMC traces. Several
parameters are strongly correlated. For instance, the energy normalization ``energy norm'' and ``$\sigma_{\rm sat}$'' are strongly correlated in that one can have less
saturation of the cross section if the energy normalization is turned down. There is also
a strong correlation between ``Init. Flow'' and ``W.N./Sat. frac.''. One can compensate for
less initial flow if the saturation formula is more heavily used than the wounded nucleon
formula. Again, this is expected because the wounded nucleon parameterization leads to
less spatial anisotropy and a somewhat more diffuse initial state.

\begin{figure}
\centerline{\includegraphics[width=0.45\textwidth]{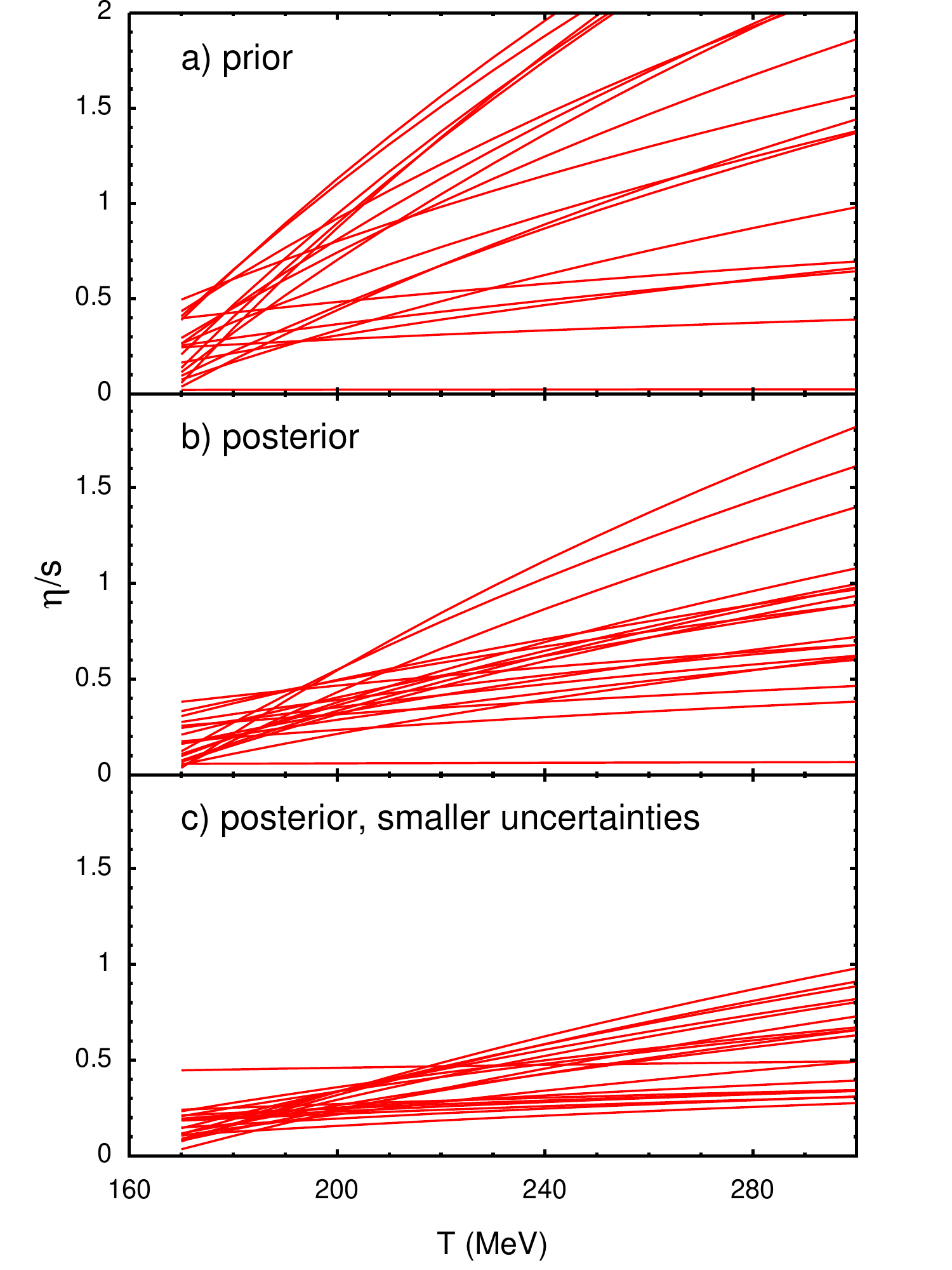}}
\caption{\label{fig:svsT}(color online)
Twenty random points in parameter space were taken from the prior (upper panel) and posterior (lower panel) distributions. The temperature dependence of the viscosity to entropy ratios is clearly constrained by the statistical comparison with data, though the posterior distribution still covers a large variation.
}
\end{figure}

The inferred viscosity is clearly correlated with the weighting between the wounded
nucleon and saturation parameterizations, as expected from the arguments in
\cite{Drescher:2007cd}. The two viscous parameters are also correlated with one another as
expected. One can compensate for a very low viscosity at $T=170$ MeV by having the
viscosity rise quickly with temperature. Figure \ref{fig:svsT} shows the viscosity to
entropy ratio as a function of temperature corresponding to the 20 random samples from the
both the prior and posterior distributions. Higher values of the temperature dependence
$\alpha$ are increasingly unlikely for higher values of $\eta/s|_{T_c}$.

\begin{figure*}[p]
\centerline{\includegraphics[width=\textwidth]{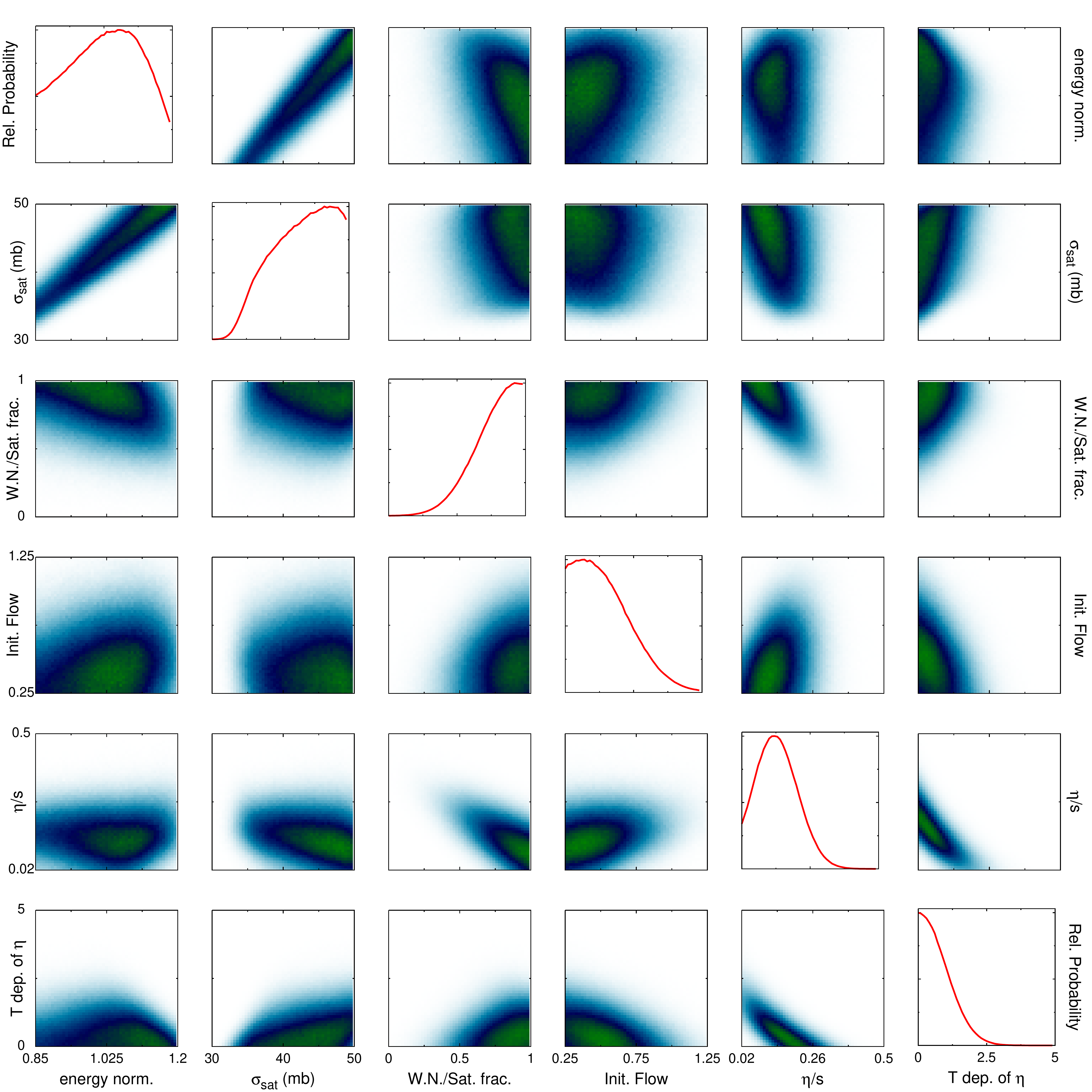}}
\caption{\label{fig:gpmcmc_halfsigma}(color online) The same as in Fig. \ref{fig:gpmcmc}, except that the more optimistic set of uncertainties from Table \ref{table:pcaobservables} were used in the analysis. By halving the uncertainties the widths of the distributions are noticeably narrower, but not by a factor of two. }
\end{figure*}

The width of the distributions in Fig. \ref{fig:gpmcmc} are influenced by the choice of uncertainties. As will be discussed in the next section, this choice is currently dominated by our lack of knowledge of how strongly missing components in the physics might affect the observables. Future study may greatly reduce these systematic theory uncertainties, or at least better quantify them. For now, we use the rather ad-hoc choices. To understand the degree to which these choices affect the posterior distribution, the statistical analysis was repeated with all uncertainties reduced by a factor of two, and are shown in Fig. \ref{fig:gpmcmc_halfsigma}. The widths of the posterior distributions do not necessarily reduce by a factor of two, because some of the widths are the result of projecting narrow distributions in higher dimension onto the one-or-two dimensional plots in Fig. \ref{fig:gpmcmc_halfsigma}. Even though the widths of the projected posterior distribution do not reduce by a factor of two, the narrowing is significant and suggests that a detailed analysis of model uncertainties would be helpful.

\section{Summary and Outlook}

Two principal conclusions can be taken from this study. First, the data from relativistic
heavy ion collisions is well suited to a multi-dimensional analysis featuring model
emulators. The response of the data to model parameters appears sufficiently smooth to
warrant simple interpolation of a few principal components. Only a half dozen parameters
were varied in this study, and only a limited number of observables were
considered. Nonetheless, the procedure should readily scale to larger numbers of parameters
and larger data sets. The successes of the emulator in reproducing model output, and of the
MCMC procedure in identifying likely regions in parameter space provide hope that the
field can produce quantitative statements concerning the bulk properties and dynamics of
the matter formed in heavy ion collisions. The second conclusion centers on the extracted
parameters. Although the ranges are subject to change given expected improvements in both
data and modeling, the ranges of parameters and correlations shown in
Fig. \ref{fig:gpmcmc} are remarkably close to expectations from less rigorous searches.

The statistical procedures applied here represent a significant improvement to the
state-of-the-art for comparisons of data and models in the field of relativistic heavy ion
physics. Previously, parameters were varied either individually, or in small
groups. Figure \ref{fig:gpcheck} demonstrates the success of using emulators for this
problem. Most importantly, the emulator techniques should scale well with increased data
and increased number of parameters. Ultimately, the number of simultaneously varied
parameters might increase to around 20, with the expectation that many, or perhaps most,
of these parameters will not be significantly constrained by the data. Additional
parameters to which the model is insensitive does not increase the need for additional
runs, as long as the parameters are varied in such a way that those parameters that are
important are well sampled. It is expected that the additional parameters would decrease
the efficiency with which the critical parameters are sampled, but that the additional
number of model runs would not make the problem intractable. Adding more data should,
hopefully, increase the number of principal components extracted from the model runs by
providing additional discriminating power. Once the model calculations have been
performed, the numerical cost of the statistical analysis performed here was negligible,
and adding more principal components should not cause any problems. Thus, the results of
this study are promising, and encourage extending the scope to much larger data sets and
more realistic models.

Although the models used here represented the current state of the art, several
improvements are necessary before firm quantitative conclusions can be extracted. The
following improvements require significant development, but are all tractable.
\begin{itemize}\itemsep=0pt
\item A flexible equation of state. For this study, the equation of state was fixed. There
  is some uncertainty involved with lattice calculations that should be accounted for with
  a variable equation of state. Additionally, it is of interest to address whether the
  equation of state is constrained by experiment alone, i.e. without relying on lattice
  calculations.
\item The bulk viscosity was set to zero here. Near $T_c$, the system is undergoes a rapid
  change in microscopic structure, and the system may lose equilibrium. As the system
  returns to equilibrium, entropy is generated. If the departure from equilibrium is
  small, the effect can be accounted for by adding a bulk viscosity
  \cite{Paech:2006st,Karsch:2007jc}. If the departure is large, other approaches are
  possible, such as dynamically solving for the mean fields \cite{Paech:2003fe}.
\item The initial chemical composition of the hadronic phase was set by the assumption of chemical
  equilibrium when the system reached a temperature of 170 MeV. The chemical evolution can
  be improved by incorporating more inelastic processes like baryon-antibaryon annihilation \cite{Pan:2012ne,Steinheimer:2012rd}. Further, the assumption of perfect equilibrium at
  a fixed temperature should be relaxed by parameterizing non-equilibrium effects.
\item Although collisions produce many thousands of particles, the initial collision
  involves only on the order of 100 nucleons. The finite number of original scattering
  centers leads to lumpy initial conditions, unlike the smooth initial conditions used
  here. If the model used here were improved to incorporate initial fluctuations, $v_2$
  would be more realistically modeled, and it would make it possible to consider
  fluctuations of the flow encoded in higher harmonics, i.e., $v_3,v_4\ldots$. Additionally, experiments analyze elliptic flow with several methods that also vary at the 10\% level. Since the methods also differ due to finite numbers of particles, once the fluctuating initial conditions are better understood, a decision needs to be made as to which method of experimentally determining $v_2$ is most appropriate for comparison with models. 
  \cite{Schenke:2012wb}.
\item Although the model used here can incorporate three-dimensional flow, for this study
  the calculations were performed with the Bjorken ansatz. This approximation is
  reasonable for collisions at 200$A$ GeV or higher \cite{Vredevoogd:2012ui}, but full
  three-dimensional calculations are needed for lower energy, or for observables away from
  mid-rapidity. Using low-energy or non-mid-rapidity data will also necessitate a more
  complex parameterization of the initial state.
\item During the hadronic phase, pionic phase space becomes highly filled at low
  $p_t$. This affects spectra at the 10\% level, which is neither crucial or
  non-negligible. Hadronic cascades can incorporate such effects by adding $(1+f)$ phase
  space enhancements to scatterings. This increases the numerical cost of the modeling at
  the factor of two level.
\end{itemize}
Most of the improvements listed above would be accompanied by an increase in the number of
parameters. For example, varying the equation of state will involve the addition of a few
parameters. Since the bulk viscosity is not well determined by lattice calculations, both
it and its temperature dependence require parameterization. Non-equilibrium chemistry can
be parameterized by adding fugacities for the initial state. Initial
conditions away from mid-rapidity, or for collisions at lower energy, could necessitate a
half dozen new parameters. At lower energy, the dependence of the equation of state on
baryon number is unknown and requires parameterization. The initial conditions at the LHC
require additional parameters to encapsulate the beam energy dependence of the initial
density and flow profiles. The lumpiness of the initial state involves setting a
transverse size of the fluctuations. It is easy to imagine future analyses involving on
the order of 20 parameters or more. 

Once the uncertainties of the models are better understood, or at least parameterized, the experimental uncertainties should dominate the expression of the uncertainties. At this point the statement of uncertainties should be revisited. Instead of the rather ad-hoc choices used here, the uncertainty of each observable needs to be expressed, a process that will require collaboration with the experimental community. Even if the uncertainties used to calculate the likelihood are purely experimental, the theoretical uncertainties encapsulated by variable parameters might still provide the dominant source of the width of the final likelihood distributions. For example, the width of the $\eta/s$ distribution might turn out to be largely determined by the correlation of the $\eta/s$ parameter with an additional poorly constrained model parameter.

This study has also largely ignored the question of which observables are mostly affected by a given parameter, or the similar question of which observable, or linear combination of observables, are most responsible for the constraining a given parameter. Given that some observables are more constrained by the variability within the model space than by the actual comparison to data, and that some of the dependencies are non-linear, robust criteria need to be developed to address these questions. This will be the focus of future studies.

The amount of available data for analyses such as these has swelled in the past few
years. The beam-energy scan at RHIC will provide all the observables analyzed here at a half
dozen more energies. Additionally, Cu+Cu, Cu+Au and U+U collisions have been measured at
RHIC. Additionally, results from Pb+Pb collisions at the LHC have now been analyzed and
published. Finally, higher flow harmonics, $v_2,\ldots, v_n$, have also become available.

Expanding the scope of the analysis to a larger range of beam energies and to include
initial state fluctuations could increase the numerical cost of the calculations by two
orders of magnitude. In the present study one processing core could perform a full model
run for one point in parameter space in approximately one day. This would increase to
being on the order of several weeks, or one month, if the beam energy scan, LHC data, and
initial-state fluctuations were included. If the number of points sampled in parameter
space were increased to a few thousand to better account for the larger number of
parameters, the project would remain tractable, but would clearly require significant
allocation of resources. The success and scalability of the methods presented here suggest
that such an effort could transform heavy ion physics into a more rigorously
quantitative science.

\begin{acknowledgments}
This work was supported by the National Science Foundation's Cyber-Enabled Discovery and Innovation Program through grant NSF-0941373 and by the Department of Energy Office of Science through grant number DE-FG02-03ER41259. RW received additional support from the National Science Foundation Division of Mathematical Sciences through grant NSF-DMS-1228317 and from the National Aeronautics and Space Administration, grant no. NNX09AK60G
\end{acknowledgments}


\end{document}